\documentclass[usenatbib]{mnras}

\usepackage{newtxtext,newtxmath}
\usepackage[T1]{fontenc}

\usepackage{graphicx}	
\usepackage{amsmath}
\usepackage{xcolor}
\usepackage{array}

\newcolumntype{P}[1]{>{\centering\arraybackslash}p{#1}}

\newcommand{\LC}{{\rm LC}}
\newcommand{\Yp}{{\rm Y}}

\newcommand{\jcop}{JCP}
\newcommand{\jpp}{JPP}
\newcommand{\cqg}{CQG}
\newcommand{\cpc}{Comput. Phys. Commun.}

\title[Diffusivity in FFE magnetosphere simulations]{Diffusivity in force-free simulations of global magnetospheres}

\author[Mahlmann and Aloy]{
J. F. Mahlmann,$^{1}$\thanks{E-mail: mahlmann@princeton.edu}
M. A. Aloy,$^{2,3}$
\\
$^{1}$Department of Astrophysical Sciences, Princeton University, Princeton, NJ 08544, USA\\
$^{2}$Departament d’Astronomia i Astrofísica, Universitat de València, 46100 Burjassot (Valencia), Spain\\
$^{3}$Observatori Astronòmic, Universitat de València, 46980 Paterna (València), Spain
}

\date{Accepted 2021 September 27. Received 2021 August 02}

\pubyear{2021}

\begin{document}

\label{firstpage}
\pagerange{\pageref{firstpage}--\pageref{lastpage}}
\maketitle

\begin{abstract}
Assuming that the numerical diffusivity triggered by violations of the force-free electrodynamics constraints is a proxy for the physical resistivity, we examine its impact on the overall dynamics of force-free aligned pulsar magnetospheres endowed with an equatorial current sheet. We assess the constraint violations as a diffusivity source. The effects of modifications on electric fields used to restore force-free conditions are not confined to the equatorial current sheet, but modify the magnetospheric dynamics on timescales shorter than the pulsar rotational period. These corrections propagate especially via a channel that was unexplored, namely, changes induced to the electric charge density, $\rho$. We quantify the global consequences of diffusivity by comparing different techniques to model $\rho$. By default, we combine a conservative $\rho$-evolution with hyperbolic/parabolic cleaning of inaccuracies in the Maxwell equations. As an alternative, we enforce a constrained evolution where $\rho$ is directly computed as the electric field divergence. The conservative approach reduces the Poynting flux dissipated in the equatorial current sheet by an order of magnitude, along with an increase of the pulsar luminosity driven by a shift of the Y-point location. The luminosity changes according to $L_\Yp\propto \alpha^{0.11}$, where  $\alpha$ is the ratio of diffusion to advection timescales, controlling the amount of (numerical) diffusivity. Our models suggest interpreting the luminosity dependence on the Y-point location as differences in resistivities encountered at the equatorial current sheet. Alternatively, they could be interpreted in terms of the pair formation multiplicity, $\kappa$, smaller diffusion being consistent with $\kappa\gg 1$.

\end{abstract}


%%%%%%%%%%%%%%%%%%%%%%%%%%%%%%%%%%%%%%%%%%%%%%%%%%

%%%%%%%%%%%%%%%%% BODY OF PAPER %%%%%%%%%%%%%%%%%%

\section{Introduction} 
\label{sec:introduction}

Energy flows induced into magnetically dominated relativistic magnetospheres of compact objects are commonly modeled by numerical simulations in the force-free electrodynamics (FFE) limit. Fueled by the track record of observations in the era of multi-messenger astrophysics, current targets for such simulations include the magnetospheres of rapidly spinning black holes, spiraling neutron stars, magnetars, and pulsars. The tenuous, magnetically dominated atmosphere (magnetosphere) of pulsars is an active field of scientific interest. They fascinate both observers \citep[e.g.][]{Lorimer_1995MNRAS.273..411,Ransom_2005Sci...307..892,Abdo_2013ApJS..208...17,Jankowski_2018MNRAS.473.4436} and theorists \citep[e.g.][]{Kennel1984,Lyubarskii1996,Contopoulos1999,Contopoulos_2019MNRAS.482L..50,Goodwin_2004MNRAS.349..213,Timokhin2006,Timokhin_2013MNRAS.429...20, Petri_2020Univ....6...15}. With the remarkable progress in scientific computing, their rotating magnetosphere has captured designers of numerical methods that integrate FFE and magnetohydrodynamics (MHD) with ever improving accuracy \citep[e.g.][]{Komissarov_2006MNRAS.367...19,Spitkovsky2006,Tchekhovskoy2013,Parfrey2017,Carrasco_2020PhRvD.101f3017}. 
%More 
Recently, particle-in-cell (PIC) simulations were able to resolve a broad range of scale separations and allow for unprecedented insight into the microphysics of pulsar magnetospheres across the global scale  \citep[][]{Cerutti_2015MNRAS.448..606,Philippov2015,Guepin2020,Kalapotharakos_2018ApJ...857...44, Philippov_2018ApJ...855...94}. In this fascinating flurry of outcomes,
only few references scrutinised whether the results from \emph{ideal} plasma simulations are the best possible model for the pulsar magnetosphere that contains an inherently \emph{non-ideal} region, namely the equatorial current sheet (ECS) beyond the closed zone \citep{Contopoulos2016,Contopoulos_2019MNRAS.482L..50, Contopoulos2020}. Here, we study with rigorous technical depth how this non-ideal region can affect the global dynamics of the force-free aligned rotator magnetosphere - effectively serving as a blueprint for force-free magnetospheres of other compact objects. 

\emph{Ideal} FFE evolve Maxwell's equations for the electromagnetic fields $\mathbf{E}$ and $\mathbf{B}$ while rigorously maintaining the force-free conditions $\mathbf{E}\cdot\mathbf{B}=0$ (equivalent to the no Ohmic heating condition $\mathbf{E}\cdot \mathbf{j}=0$, where $\mathbf{j}$ is the electric current) and $\mathbf{E}^2-\mathbf{B}^2<0$ (magnetic dominance). \emph{Non-ideal} force-free fields are those fields that allow for perturbations to the condition $\mathbf{E}\cdot\mathbf{B}=0$ by non-negligible electric fields $E_\parallel$ along the magnetic field $\mathbf{B}$ \citep[e.g.][]{Lyutikov_2003MNRAS.346..540}. Such fields were used in the literature to induce the concept of resistivity into FFE, either by specifically designed driving currents \citep[e.g.,][]{Alic2012} or by self-consistently modelled alterations to the current of ideal FFE \citep{Komissarov2004,Parfrey2017}. 

In numerical models of FFE, it is common to enforce the preservation of the $\mathbf{E}\cdot\mathbf{B}=0$ and $\mathbf{E}^2-\mathbf{B}^2<0$ conditions algebraically by resetting the electric field \emph{instantaneously} wherever they are not fulfilled \citep[e.g.][]{Palenzuela2010,Mahlmann2020}.
Exploiting the specifics of our numerical methodology \citep{Mahlmann2020b}, which combines the instantaneous algebraic extraction of all non-ideal electric fields while evolving the charge continuity equation, we identified another mechanism that adds diffusivity to an ideal FFE scheme  \citep{Mahlmann2020c}. Namely, the misalignment of electric fields $\mathbf{E}$ and charge density $\rho$ can significantly alter an FFE evolution. In this numerical survey, we further develop the findings obtained with an idealised setup that triggers tearing modes in force-free current sheets \citep{Mahlmann2020c} to the astrophysical relevant scenario of pulsar magnetospheres.

In the not uncontroversial realm of magnetospheric simulations in the force-free limit, we announced in \citet{Mahlmann2020b} that ambiguities in the standard reference of the force-free aligned rotator required further attention. In fact, \citet{Contopoulos2016} pointed out these ambiguities (\emph{trade secrets}) that arise when simulating magnetospheres with non-ideal regions, such as the pulsar and Wald magnetospheres, in the ideal plasma limit, as one finds in ideal MHD and FFE.  A sensitivity for the dazzling amount of calibration that time-dependent numerical simulations require is rarely transmitted along with the visually appealing results themselves. This manuscript is an effort to bring transparency to the modeling of one astrophysical scenario that crosses the constraints set by FFE. It aims at enabling the reader to ask crucial questions when evaluating results from the simulations of magnetospheres and intends to place some landmarks for the development of future hybrid methods that are not restricted to the ideal regime.

This manuscript is organised as follows. In Sect.~\ref{sec:methodology}, we review the employed numerical methodology as well as the pulsar magnetosphere initial data (Sect.~\ref{sec:simulationsetup}). Sect.~\ref{sec:forcefreealigned} presents the outcome of the conducted simulations of a force-free aligned rotator. The results are grouped by different topics. First, we examine the dependency of the luminosity at the light cylinder (LC) on the method employed to preserve the consistency between the charge distribution and the currents (Sect.~\ref{sec:lcluminosity}). Section~\ref{sec:econservation} analyses the conservation of energy beyond the light cylinder. An array of ancillary high-resolution models yields additional insights into the subtleties of ideal FFE simulations in Sect.~\ref{sec:tradesecrets}. Specifically, we assess the role of the magnetic dominance condition (Sect.~\ref{sec:focuseddominance}), compare algebraic corrections of force-free violations to driving currents (Sect.~\ref{sec:drivingfocus}), and study the effect of resistivity models beyond the light cylinder (Sect.~\ref{sec:diffusivityfocus}). The discussion of Sect.~\ref{sec:discussion} includes views on the propagation of force-free violations (Sect.~\ref{sec:nonidealFF}), the diffusive time scales set by the employed hyperbolic/parabolic cleaning (Sect.~\ref{sec:cleaningscales}), and a general picture of diffusivity in force-free magnetospheres (Sect.~\ref{sec:diffusivitydiscuss}). We conclude this survey by summarizing the main takeaways of the presented results in Sect.~\ref{sec:conclusion}.

\section{Methodology}
\label{sec:methodology}

The aligned rotator problem has been studied vastly throughout the last 20 years and now appears to be a well-established test case for FFE codes. Even if it is likely a simplification of the more complex problem of a magnetic dipole that is misaligned with respect to the rotational axis of the pulsar, it still contains an element of special relevance for the overall problem, namely, the ECS. We approach this investigation by means of time-dependent FFE simulations performed with the numerical code presented in \cite{Mahlmann2020b,Mahlmann2020c}. Our method is an enhanced high-order conservative realization of FFE as introduced by \citet{Komissarov2004} and vastly benefits from the \textsc{Carpet} driver \citep{Goodale2002a,Schnetter2004} and its extension to spherical coordinates \citep{Mewes2018,Mewes2020} supported by the infrastructure of the \textsc{Einstein Toolkit}\footnote{\url{http://www.einsteintoolkit.org}}.

The scheme that we introduced in \cite{Mahlmann2020b} has since been compared to another force-free MHD method \citep[\textsc{BHAC},][]{Ripperda2019,Ripperda2019a} in the context of Alfvén wave interactions in the highly magnetised limit \citep{Ripperda2021}. It achieved a striking convergence of results across different methods.
Also in the context of the aligned force-free rotator \citep[cf. Sect. 5.2 in][]{Mahlmann2020b}, our FFE method reproduced the main characteristics of the pulsar magnetosphere: co-rotating magnetic field lines, a force-free closed zone in the wake of the Y-point, and an ECS. 
However, \cite{Mahlmann2020b} observed a shift of the Y-point away from the light cylinder and a Poynting flux that was above the value which is treated as an established reference throughout the literature \citep[e.g.,][]{Spitkovsky2006, Tchekhovskoy2013, Etienne2017}. In the following subsections we briefly introduce the governing equations for the problem at hand, and the methods used to integrate them. We also provide the detailed numerical setup in Sect.~\ref{sec:numericalmethod}.

\subsection{Numerical method}
\label{sec:numericalmethod}
We employ the force-free scheme presented in \citet{Mahlmann2020b} to conduct 2D simulations of pulsar magnetosphere on a flat background without spacetime curvature \citep[cf. Sect. 5.2 in][]{Mahlmann2020b}. Ignoring general relativistic effects \citep[negligible for the global dynamics of the pulsar magnetosphere, especially far away from the neutron star surface, though very relevant, especially frame-dragging, for driving efficient pair production;][]{Philippov_2015ApJ...815L..19, Belyaev_2016ApJ...830..119, Gralla_2016ApJ...833..258, Philippov_2018ApJ...855...94}, the equations of FFE are the set of partial differential equations formed by the Maxwell equations together with the corresponding solenoidal constraint
($\nabla\cdot\mathbf{B}=0$) and the charge density, $\rho$ expressed as $\rho=\nabla\cdot\mathbf{E}$. These two equations are integrated in our code by employing the so-called hyperbolic/parabolic cleaning method \citep{Dedner2002,Komissarov2007MNRAS.382..995}, so that the former elliptic constrains become hyperbolic equations and form an \emph{augmented} system of equations:
\begin{align}
\partial_t \mathbf{B}    &= -\nabla \times \mathbf{E}-c_\Psi^2\nabla\Psi\\
\partial_t \mathbf{E}    &= \nabla \times \mathbf{B}+c_\Phi^2\nabla\Phi-\mathbf{j} \label{eq:Efield}\\
\partial_t \Psi &= -\nabla\cdot \mathbf{B}-\kappa_\Psi \Psi \label{eq:Psi}\\
\partial_t \Phi &= \nabla\cdot \mathbf{E}- \rho-\kappa_\Phi \Phi \label{eq:Phi}\\
\partial_t \rho &= -\nabla\cdot \mathbf{j} \label{eq:chargeconservation}
\end{align}
 Our base numerical scheme employs the scalar potentials $\Phi$ and $\Psi$ to handle (numerical) errors to the constraints $\nabla\cdot\mathbf{E}=\rho$ and $\nabla \cdot\mathbf{B}=0$, respectively \citep[cf.][]{Komissarov_2006MNRAS.367...19,Mignone2010}. The action of the scalar potentials is controlled by a combination of damping constants, $\kappa_\Psi$ and $\kappa_\Phi$, as well as a set of advection speeds $c_\Psi$ and $c_\Phi$. We will refer to our default scheme as the charge conservative (CC) method, since it involves the charge conservation equation \eqref{eq:chargeconservation}, and guarantees that the charge distribution is consitent with the currents in the domain \citep{Komissarov_etal_2007MNRAS.374..415}. 

A principal ingredient of our methodology is the current density as observed by the normal observer \citep[cf.][]{Mahlmann2020b,Mahlmann2020c}. It naturally splits into components 
perpendicular and parallel to the magnetic field three-vector ($\mathbf{j}=\mathbf{j}_\perp + \mathbf{j}_\parallel$):
\begin{align}
	\mathbf{j}_{\perp} &= \rho \frac{\mathbf{E} \times \mathbf{B}}{\mathbf{B}^2}, \label{eq:FFResCurrentPerpendicular1}\\
	\mathbf{j}_\parallel &= %\frac{1}{1 + \kappa_I\eta} 
	\frac{  \mathbf{B} \cdot(\nabla\times\mathbf{B})  - 
		\mathbf{E}\cdot(\nabla\times\mathbf{E}) + \kappa_I \: \mathbf{B}\cdot \mathbf{E}}{(1 + \kappa_I\eta)\,\mathbf{B}^2} \:\mathbf{B} 
	\label{eq:FFResCurrentPerpendicular}.
\end{align}
Here, $\kappa_I$ is the decay rate driving the electric field toward its target value $\mathbf{E}\rightarrow \eta\mathbf{j}$, and $\eta$ is a dissipation coefficient for the electric field that is parallel to the current. In the following sections, we specify whenever we extend the current of \emph{ideal} FFE ($\eta=\kappa_I=0$).

In a variation of the default numerical scheme, we ignore the charge conservation equation \eqref{eq:chargeconservation} and 
compute the charge density appearing in the source terms by imposing $\rho=\nabla\cdot\mathbf{E}$. Specifically, the divergence of the electric field is computed from the cell-centered (volume averaged) electric field values in Eqs.~(\ref{eq:Efield}) and~(\ref{eq:FFResCurrentPerpendicular1}). We maintain the hyperbolic equation for the scalar potential $\Phi$ \eqref{eq:Phi} in order to dissipate and transport away any misalignment between charges and currents and preserve the charge conservation equation (\ref{eq:chargeconservation}) up to truncation error. We note that the $\nabla\cdot\mathbf{E}$ term appearing in Eq.~\eqref{eq:Phi} is computed as a numerical flux for the temporal update of $\Phi$ and, as such, it is obtained from the inter-cell values of the electric field. These interface values are monotonically reconstructed from cell-centred volume averaged values of the electric field. Hereafter, we will refer to this variation of the base scheme as the local charge reconstruction (LCR) method. We note that we employ a finite volume method where none of the variables is staggered off the cell centers \citep[differently from, e.g.][]{Spitkovsky2006, Mignone_2019MNRAS.486.4252}. Thus, the evaluation of $\nabla\cdot\mathbf{E}$ at each numerical cell is performed by employing a fourth order finite difference approximation based on a suitable number of neighboring cell values of $\mathbf{E}$ \citep[with an stencil similar to that used for the evaluation of $\mathbf{j}_\parallel$,][cf. Sect.~3.4]{Mahlmann2020}.

Finally, we also considered a second variation of the default numerical scheme that combines the two previous strategies into a \emph{hybrid} method (HCC hereafter). In the HCC algorithm we restrict the use of the LCR method to places where
the magnetic dominance condition is violated during any sub-step of the time integration, and the CC method is applied elsewhere. In practice, for the specific context of the aligned rotator magnetosphere, this limits the application of the LCR method to numerical cells affected by the ECS.

\subsection{Simulation setup}
\label{sec:simulationsetup}

As it is a common practice in the field, we employ a non-rotating dipole magnetosphere as initial data \citep[see Sect.~5.1 in][]{Mahlmann2020b}. The initially purely poloidal magnetic flux in the magnetosphere is then 
\begin{align}
    \mathbf{B}&=\mu\left(\frac{2\cos\theta}{r^3},\frac{\sin\theta}{r^4},0\right),\\
    \left|\mathbf{B}\right|&\equiv B_{\rm d}\left(r,\theta\right)=\frac{\mu}{r^3}\left[3\cos^2\theta+1\right]^{1/2},
    \label{eq:Bdipole}
\end{align}
where we scale the magnetic moment by the stellar radius, $\mu = r_*^{3/2}$ and the vector components are expressed in the orthogonal spherical basis for the coordinates $\{r,\theta,\phi\}$. An axisymmetric rotation is instantaneously switched on across the stellar surface, so that there is a transient period during which a torsional Alfvén wave propagates outwards throughout the magnetosphere. After this transient period, an \emph{approximately} steady state is reached
%We initialise a magnetic dipole field
in the domain of extensions $r\times\theta=\left[r_*,751r_*\right]\times\left[0,\pi\right]$. Here, we use the stellar radius $r_*=13.67\,$km \citep[cf.][]{Mahlmann2019}. We evolve the magnetosphere in time for $t=8.35t_{\rm p}$, and $t=30.24t_{\rm p}$ for selected cases, where $t_{\rm p}=2\pi/\Omega_{\rm p}$ is the time of one pulsar revolution, and we choose $\Omega_{\rm p}\approx 646\,$Hz (equivalent to $\Omega_{\rm p}=0.02$ in the units of our numerical code). With this choice, the LC is located at distance $r_\LC\equiv c/\Omega_{\rm p}=5r_*$.

The study presented here evaluates numerical convergence for increasing resolution carefully, using combinations of the radial spacing $\Delta r=r_*/N_r$, $N_r\in\left[32,64,128\right]$, and angular spacing $\Delta \theta=\pi/N_\theta$, $N_\theta\in\left[100,200,400\right]$, respectively. The light-cylinder region is, thus, covered by a minimum of $N_{\rm LC}\in\left[160,320,640\right]$ radial mesh cells, while the total number of radial grid points exceeds this number by a large factor. We note that the outer boundary is sufficiently far away from the central object to avoid any feedback on the star itself. The inner boundary, located at the stellar surface, requires careful attention. To first order, we follow the treatment suggested by \citet{Parfrey2012}. We set the value of $B^r$ not only at the surface, but also within a thin layer $r_{\rm in} \le r\le r_*$. The remaining magnetic field components are treated differently. Namely, they are driven to their dipole values (Eq.~\ref{eq:Bdipole}) at some distance from the surface. In a thin boundary layer consisting of at least as many cells to be covered twice by the stencil of the chosen spatial reconstruction, $B^\theta$ and $B^\phi$ are evolved with the Maxwell equations. Deep inside the interior boundary, all deviations from the initial dipole fields are exponentially damped. Within the whole boundary layer, the electric field is obtained by assuming that it is purely inductive, i.e. \citep[cf.][]{Parfrey2012},
\begin{align}
    \mathbf{E}=-\left[\mathbf{\Omega}_{\rm p}\times \mathbf{r}\right] \times\mathbf{B} . 
\end{align}

This rather complicated boundary is needed to prevent spurious numerical behaviors at the stellar surface, which happen if all the magnetic field components are held fixed at $r_{\rm in} \le r \le r_*$. Likewise, this boundary acts equivalently to a conducting boundary in as much as it preserves the continuity of the electric field components parallel to the surface and the magnetic field perpendicular to it (i.e., no jumps in $B^r$, $E^\theta$ or $E^\phi$ develop at the stellar surface). It allows to relax the initial values of the electromagnetic field at the stellar surface to the approximate equilibrium values attained after a few rotational periods.
The electric charge inside the pulsar boundary layer is calculated in every sub-step of the time integration via $\rho=\nabla\cdot\mathbf{E}$. To ensure consistency at the inner boundary, we set $\Phi=0$ for $r<r_*$, while evolving $\Psi$ freely to evacuate errors to the solenoidal constraint on the magnetic field $\mathbf{B}$.

The evaluation of results presented throughout the following sections relies on the comparison of dimensionless quantities. Therefore, we normalise magnetic fields by the polar magnetic field $B_0=B_{\rm d}\left(r_*,0\right)=2\mu/r_*^3$. The charge density is normalised to the Goldreich-Julian charge density \citep{Goldreich_Julian_1969ApJ...157..869} at the polar cap, namely,
\begin{align}
    \rho_0=\rho_{\rm GJ}\left(r_*,0\right)=-\frac{2\Omega_{\rm p} B_0}{c},
\end{align}
where $c$ is the speed of light ($c=1$ in the units of our code). Equally, electromagnetic currents will be normalised by the reference current density $j_0=|\rho_0c|$.

\begin{table}
\centering
\caption{Properties of the simulations corresponding to models described in Sect.~\ref{sec:forcefreealigned}. We provide the label of the respective model, the   number of zones per stellar radius (in total, we have $\ge 750\times N_r$ radial zones) and in the interval $[0,\pi]$, the strategies to model the electric charge, the $\alpha$ parameter, the Y-point location, the luminosity at the LC, its relative decay up to a distance of $5r_{\rm LC}$, and the colatitude of the closed zone separatrix at the stellar surface. All models of this section evacuate force-free constraint violations by algebraic cropping of the electric fields.} \label{tab:mainmodels}
\begin{tabular}{P{0.2cm}P{1.2cm}P{0.4cm}P{0.5cm}P{0.5cm}P{0.7cm}P{0.8cm}P{0.4cm}}
\hline
\hline
 & $N_r\times N_\theta$ & $\rho$ & $\alpha$ & $x_0$ & $L_\Yp/L_0$ & $\Delta L/L_\Yp$ & $\theta_{\rm c}$\\
\hline
\textbf{La} & $32\times 100$ & LCR & $72$ & $0.94$ & $0.97$  & $0.36$ & $33.2^\circ$\\%$0.58$ \\
\textbf{Lb} & $32\times 100$ & CC & 0.03 & $0.92$ & $1.03$ & $0.30$ & $33.8^\circ$\\%$0.59$ \\
\textbf{Lc} & $32\times 100$ & CC & $0.3$ & $0.88$ & $1.24$ & $0.08$ & $35.0^\circ$\\%$0.61$ \\
\textbf{Ld} & $32\times 100$ & CC & $2.9$ & $0.83$ & $1.67$ & $0.03$ & $36.7^\circ$\\%$0.64$ \\
\textbf{Le} & $32\times 100$ & CC & $9.3$ & $0.78$ & $1.95$ & $0.03$ & $37.8^\circ$\\%$0.66$ \\
\textbf{Lf} & $32\times 100$ & CC & $18$ & $0.71$ & $2.06$ & $0.05$ & $38.4^\circ$\\%$0.67$ \\
\textbf{Lg} & $32\times 100$ & CC & $37$ & $0.78$ & $2.12$ & $0.05$ & $38.4^\circ$\\%$0.67$ \\
\textbf{Lh} & $32\times 100$ & CC & $72$ & $0.77$ & $2.11$ & $0.04$ & $37.8^\circ$\\%$0.66$ \\
\textbf{Li} & $32\times 100$ & CC & $0.2$ & $0.89$ & $1.15$ & $0.10$ & $35.0^\circ$\\%$0.61$ \\
\textbf{Lj} & $32\times 100$ & CC & $0.6$ & $0.86$ & $1.45$ & $0.04$ & $36.1^\circ$\\%$0.63$ \\
\textbf{Lk} & $32\times 100$ & CC & $4.6$ & $0.79$ & $1.65$ & $0.04$ & $36.7^\circ$\\%$0.64$ \\
\textbf{Ll} & $32\times 100$ & CC & $19$ & $0.70$ & $2.21$ & $0.05$ & $38.4^\circ$\\%$0.67$ \\
\textbf{Lm} & $32\times 100$ & CC & $9.3$ & $0.78$ & $1.80$ & $0.06$ & $37.8^\circ$\\%$0.66$ \\
\textbf{Ln} & $32\times 100$ & CC & $37$ & $0.71$ & $2.28$ & $0.08$ & $39.0^\circ$\\%$0.68$ \\
\hline
\textbf{Ma} & $64\times 200$ & LCR & $36$ & $0.96$ & $1.01$ & $0.36$ & $32.7^\circ$\\%$0.57$ \\
\textbf{Mb} & $64\times 200$ & CC &  0.02 & $0.95$ & $1.06$ & $0.35$ & $33.2^\circ$\\%$0.58$ \\
\textbf{Mc} & $64\times 200$ & CC & $0.2$ & $0.91$ & $1.21$ & $0.06$ & $33.8^\circ$\\%$0.59$ \\
\textbf{Md} & $64\times 200$ & CC & $1.5$ & $0.86$ & $1.58$ & $0.03$ & $36.1^\circ$\\%$0.63$ \\
\textbf{Me} & $64\times 200$ & CC & $4.6$ & $0.80$ & $1.87$ & $0.04$ & $36.7^\circ$\\%$0.64$ \\
\textbf{Mf} & $64\times 200$ & CC & $9.3$ & $0.74$ & $1.95$ & $0.05$ & $37.2^\circ$\\%$0.65$ \\
\textbf{Mg} & $64\times 200$ & CC & $18$ & $0.76$ & $2.07$ & $0.04$ &  $37.2^\circ$\\%$0.65$ \\
\textbf{Mh} & $64\times 200$ & CC & $36$ & $0.71$ & $2.20$ & $0.06$ & $37.8^\circ$\\%$0.66$ \\
\textbf{Mi} & $64\times 200$ & CC & $0.1$ & $0.92$ & $1.19$ & $0.07$ & $34.4^\circ$\\%$0.60$ \\
\textbf{Mj} & $64\times 200$ & CC & $0.3$ & $0.91$ & $1.43$ & $0.03$ & $35.0^\circ$\\%$0.61$ \\
\textbf{Mk} & $64\times 200$ & CC & $2.3$ & $0.87$ & $1.56$ & $0.02$ & $35.5^\circ$\\%$0.62$ \\
\textbf{Ml} & $64\times 200$ & CC & $9.2$ & $0.77$ & $2.17$ & $0.04$ & $37.8^\circ$\\%$0.66$ \\
\textbf{Mm} & $64\times 200$ & CC & $4.6$ & $0.83$ & $1.72$ & $0.04$ & $36.1^\circ$\\%$0.63$ \\
\textbf{Mn} & $64\times 200$ & CC & $19$ & $0.65$ & $2.37$ & $0.07$ & $39.0^\circ$\\%$0.68$ \\
\hline
\textbf{Ha} & $128\times 400$ & LCR & $18$ & $0.99$ & $0.96$ & $0.32$ & $31.5^\circ$\\%$0.55$ \\
\textbf{Hb} & $128\times 400$ & CC & $0.1$ & $0.62$ & $1.17$ & $0.10$ & $33.8^\circ$\\%$0.59$ \\
\textbf{Hc} & $128\times 400$ & CC & $0.7$ & $0.81$ & $1.53$ & $0.01$ & $34.4^\circ$\\%$0.60$ \\
\textbf{Hd} & $128\times 400$ & CC & $2.3$ & $0.72$ & $1.83$ & $0.03$ & $35.5^\circ$\\%$0.62$ \\
\textbf{He} & $128\times 400$ & CC & $4.6$ & $0.68$ & $2.04$ & $0.04$ & $36.7^\circ$\\%$0.64$ \\
\textbf{Hf} & $128\times 400$ & CC & $9.2$ & $0.51$ & $2.24$ & $0.23$ & $36.7^\circ$\\%$0.64$ \\
\textbf{Hg} & $128\times 400$ & CC & $18$ & $0.41$ & $2.07$ & $0.21$ & $37.8^\circ$\\%$0.66$ \\
\textbf{Hh} & $128\times 400$ & CC & $0.4$ & $0.90$ & $1.37$ & $<0.01$ & $34.4^\circ$\\%$0.60$ \\
\textbf{Hi} & $128\times 400$ & CC & $1.5$ & $0.64$ & $1.91$ & $0.06$ & $36.1^\circ$\\%$0.63$ \\
\textbf{Hj} & $128\times 400$ & CC & $1.2$ & $0.84$ & $1.50$ & $0.01$ & $34.4^\circ$\\%$0.60$ \\
\textbf{Hk} & $128\times 400$ & CC & $4.6$ & $0.52$ & $2.20$ & $0.18$ & $37.2^\circ$\\%$0.65$ \\
\textbf{Hl} & $128\times 400$ & CC & $2.3$ & $0.85$ & $1.59$ & $0.02$ & $34.4^\circ$\\%$0.60$ \\
\textbf{Hm} & $128\times 400$ & CC & $9.3$ & $0.36$ & $2.42$ & $0.35$ & $40.7^\circ$\\%$0.71$ \\
\hline
\hline
\end{tabular}
\end{table}

\section{Force-free aligned rotator}
\label{sec:forcefreealigned}

\begin{figure*}
  \centering
  \includegraphics[width=1.0\textwidth]{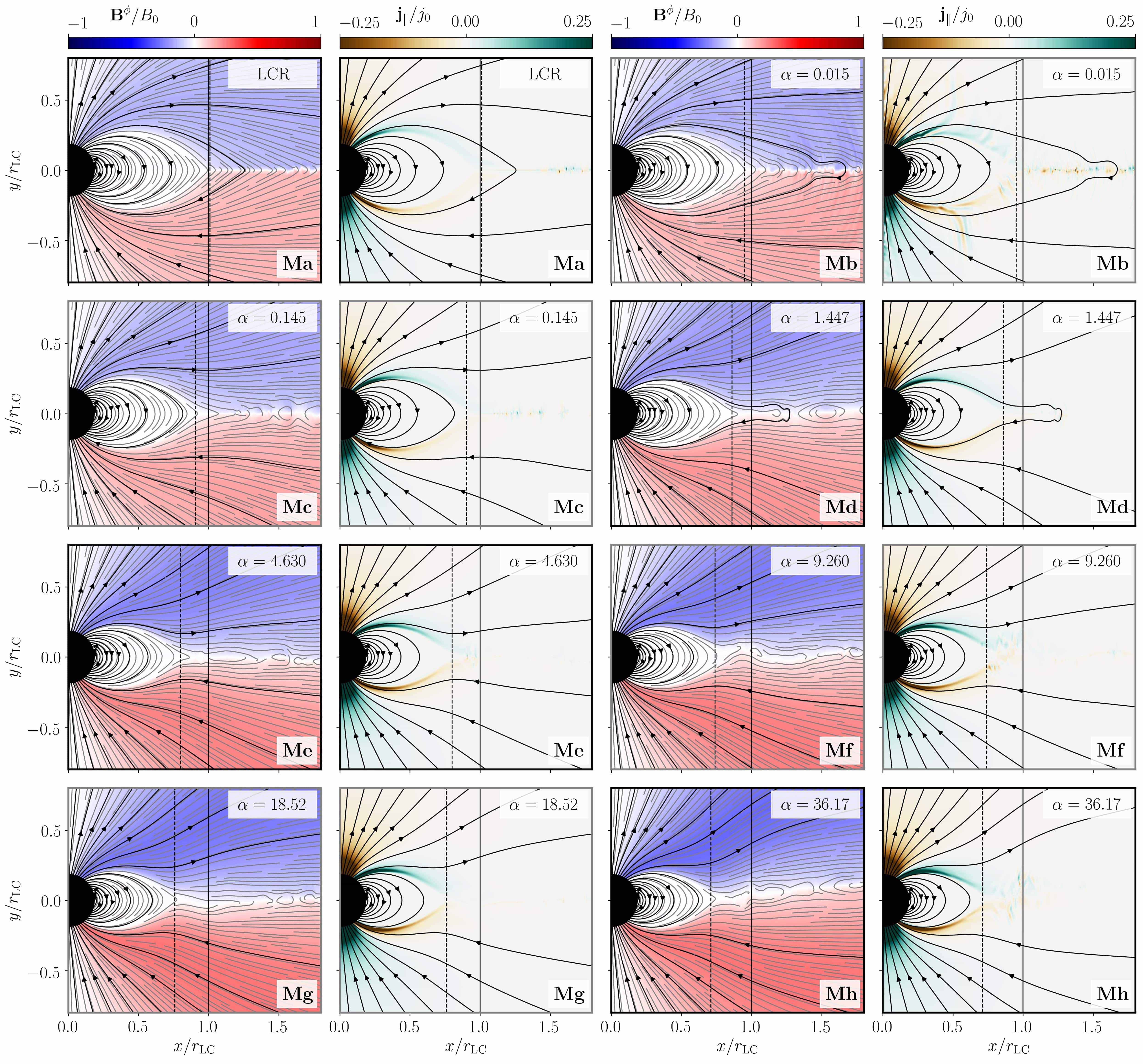}
  \vspace{-15pt}
  \caption{Magnetic field component $B^\phi$ (normalised to $B_0$) and parallel force-free current $\mathbf{j}_\parallel/j_0$ for different values of the diffusivity parameter $\alpha\in\left[0.015,0.145,1.447,4.630,9.260,18.52,36.17\right]$ in the CC and in the LCR methods for a resolution of $N_r = 64$. In all cases, we employ $c_\Phi=1$, and for the run with the LCR method, we choose the extreme value $\alpha=36.17$. Black fieldlines are seeded at the same latitude on the stellar surface and may serve as a reference points for comparability. The light cylinder position is indicated by a solid black line, and an estimate of the Y-point location by a dashed black line. 
  }
\label{fig:FLDSJDENS}
\end{figure*}

In this section, we evaluate and interpret results from an extensive array of simulations (Tab.~\ref{tab:mainmodels}). 
Specifically, Sects.~\ref{sec:econservation} and \ref{sec:lcluminosity} 
present results from 41 simulations in the \emph{ideal} FFE limit, spanning different resolutions as well as a parameter range for $\kappa_\Phi$ and $c_\Phi$. Fig.~\ref{fig:FLDSJDENS} marks the starting point - and preceding motivation - of our exploration: 
The FF equilibrium aligned rotator magnetosphere is vastly different when the LCR method is employed (top left panels), as opposed to a conservative evolution of the charge continuity equation in our CC scheme.

A useful dimensionless parameter to classify our results is
\begin{align}
    \alpha=\frac{\kappa_\Phi}{c_\Phi}\Delta h, 
    \label{eq:alphadimensionless}
\end{align}
where $\Delta h=\text{min}\left[\Delta r, r \Delta\theta, r\sin\theta\Delta\phi\right]$. We provide a detailed interpretation of $\alpha$ throughout Sect.~\ref{sec:cleaningscales}. With an increasing damping coefficient $\kappa_\Phi$ (i.e. with increasing $\alpha$) of the hyperbolic/parabolic cleaning, the Y-point separating the closed zone from the equatorial current sheet moves closer to the central object and, thus, away from the LC. Equally, the amount of reconnecting field lines through the ECS notably decreases as $\alpha$ increases. The LRC method emerges as the \emph{most diffusive} limit of our parameter exploration. In comparable force-free simulations, \citet{Komissarov_2006MNRAS.367...19} finds results that are very similar to this limit. Observing significant reconnection beyond the Y-point, it is concluded that FFE has a tendency to facilitate the \textit{development of strongly dissipative current sheets}. Such solutions  \textit{could only be relevant for magnetospheres with effective radiative cooling}. At the same time, one finds results in the literature where - without employing an explicit conservative evolution - the LCR method captures the equatorial current sheet of the pulsar magnetosphere rather well \citep{Spitkovsky2006,Etienne2017}. However, they employ different techniques to preserve the $\nabla\cdot\mathbf{E}=\rho$ and $\nabla\cdot\mathbf{B}=0$ constraints on the electromagnetic fields, based on either staggering the electromagnetic fields or employing the four-vector potential together with the energy-flux formulation of FFE \citep{McKinney2006}. We believe that scrutinizing this discrepancy up to its most finely granulated technical detail is crucial to understand the limits of FFE, how these limits affect astrophysical modeling, and how they can be overcome. 

\subsection{Luminosity at the light cylinder}
\label{sec:lcluminosity}

The Poynting flux at the LC is the most commonly cited reference value for which the modeling of FF magnetospheres of aligned rotators is consulted. \citet{Timokhin2006} presents a thorough review of the FF steady pulsar magnetosphere, and we refer to the same reference luminosity \citep[see also][]{Gruzinov_2005PhRvL..94b1101},
\begin{align}
    L_0 \approx \frac{\mu^2\Omega_p^4}{c^3}.
\end{align}
Previous results for time-dependent models show that the pulsar luminosity reached in the FF magnetosphere of an aligned pulsar \citep[with a Y-point located at the LC; note that equilibrium solutions with an inward-shifted Y-point exist][]{Goodwin_2004MNRAS.349..213,Contopoulos_2005A&A...442..579,Timokhin2006} is $L_{\rm LC}\equiv L(r=r_{\rm LC})=\left(1.0\pm 0.1\right)L_0$ \citep[e.g.][]{Komissarov_2006MNRAS.367...19,Spitkovsky2006,Tchekhovskoy2013}. In contrast, we shall argue that both the luminosity at the LC and the Y-point location depend on the (numerical) resistivity of the employed algorithm. Ultimately, this resistivity drives a slippage of the magnetic field lines at the Y-point as well as in the region of the ECS trailing it, and triggers their differential rotation in the magnetosphere. \cite{Contopoulos_2005A&A...442..579} explore the possibility that there is a differential rotational velocity of the open magnetic field lines to build solutions where the Y-point can be anywhere inside the LC. Figures~\ref{fig:Poynting} and~\ref{fig:PoyntingCSPEED} display the integrated Poynting flux through concentric spheres, as a function of distance from the central object. These plots show very different behavior beyond the LC (as we will discuss in Sect.~\ref{sec:econservation}). Indeed, we observe a transition towards the Poynting flux of the LCR method, and its dependence on distance from the star, for decreasing values of $\alpha$ in the set of data represented in Fig.~\ref{fig:Poynting}. The results shown for the CC method in Figs.~\ref{fig:Poynting} and~\ref{fig:PoyntingCSPEED} are obtained for different combinations of the cleaning parameters $\kappa_\Phi$ and $c_\Phi$ (the rest of the algorithmic elements in our numerical code are fixed). The broad range of luminosities spans between $\sim L_0$ (for the smallest value of $\kappa_\Phi=0.1$) and $\sim 2.3 L_0$ (for $\kappa_\Phi\ge 128$). There is a notable difference regarding the pulsar luminosity among different methods and within the CC method with distinct numerical parameters. As we argue below, the differences arise by the change in the location of the Y-point. Increasing $\kappa_\Phi$ and decreasing $c_\Phi$ (both changes yielding an increase of $\alpha$), thus, limiting the spread of numerical errors buffered in the cleaning potential $\Phi$, increases the luminosity at the LC. We observe a small growth of the total luminosity for $\kappa_\Phi>64$ when the numerical resolution is doubled. We will argue in Sects.~\ref{sec:tradesecrets} and \ref{sec:discussion} that this is because very large values of $\alpha$ also require very large numerical resolution to properly resolve the very fast damping of divergence cleaning errors in time. The LCR method corresponds to the limit in which errors to charge conservation spread through the domain without constraint. Hence, changes in the electric field on the ECS performed to restore the magnetic dominance condition are communicated \textit{instantaneously} all over the stencil of the discretization of the $\nabla$ operator in a single time iteration of the method (coupling as many as 12 zones around a given numerical cell, if a fourth order finite differences formula is employed). 

Figure~\ref{fig:FLDSJDENS} lucidly illustrates that the Y-point moves away from the LC with increasing values of $\alpha$. A Y-point closer to the stellar surface boosts the electromagnetic luminosity of the pulsar \citep{Timokhin2006}. The amount of open magnetic field lines increases for a decreasing dimensionless distance $x_0\equiv r_\Yp/r_\LC$ of the Y-point from the central object. This is directly related to the angular extension of the polar cap, which can be quantified by measuring the colatitude of the closed zone region (see below). In Fig.~\ref{fig:YpointLuminosity}, we present a large selection of simulated Y-point luminosities ($L_\Yp$) vs. their estimated Y-point position, excluding results that did not yield equilibrium magnetospheres for very small or very large values of $\alpha$ (see discussion in Sect.~\ref{sec:discussion}). To approximate the Y-point location, we evaluate the drift velocity along the equator and assign $x_0$ to the position where the velocity is comparable to a small parameter, which equals the grid spacing $\Delta r$ in magnitude. We employ a similar approximation technique to the evaluation of the closed zone colatitude $\theta_{\rm c}$ (Tabs.~\ref{tab:mainmodels} and~\ref{tab:ancilmodels}), with a suitably strong decay at the transition to the co-rotation region. We find that, approximately, $\sin\theta_{\rm c}\sim (r_*/r_\Yp)^{1/2} \sim 0.5x_0^{-1/2}$. Roughly in line with the findings from \citet{Timokhin2006}, we measure a correlation between the Y-point location and luminosity. For large values of $\alpha$, associated to smaller values of $x_0$, the errors in these measurements - averaged over several pulsar revolutions - become larger. These errors are likely linked to degrading numerical accuracy induced by the increased stiffness of the augmentation equations (\ref{eq:Psi}) and (\ref{eq:Phi}) for large values of $\alpha$ (see  Sect.~\ref{sec:discussion}).
We note that the correlation between $x_0$ and $L_\Yp/L_0$ found numerically approaches the theoretical relation of \citet{Timokhin2006}, in which $L_\Yp\propto x_0^{\lambda}$, with $\lambda = -2.065$, more so as the numerical resolution increases. Yet another interesting correlation derived from our results is that $L_\Yp/L_0\approx 0.57(\sin\theta_{\rm c})^{0.19}$, which supports the claim that a larger luminosity at the Y-point correlates with a larger polar cap angle. Interpreting our findings, we cautiously suggest a possibility to obtain quasi-stationary magnetospheres in which the Y-point is not located at the LC but inside it. The location of the Y-point depends on the diffusivity at the ECS. Smaller diffusivity moves the Y-point inward and increases the outgoing Poynting flux, as we show in Fig.~\ref{fig:YpointLuminosity}. 
The root of this dependency is the (magnetic fieldline) coupling between the fieldline footpoints at the stellar surface and part of the ECS (precisely, the part adjacent to the Y-point). The existence of such a region of coupling between the stellar surface and the ECS was suggested, e.g., by \citet{Contopoulos_2019MNRAS.482L..50} for stationary hybrid solutions combining FFE and a non-ideal ECS. Our time-dependent models reproduce many aspects of these equilibrium states. In the right panel of Fig.~\ref{fig:YpointLuminosity} we evaluate the specific dependency of the luminosity on the diffusion parameter $\alpha$. A power-law of the form $L_\Yp/L_0\propto \alpha^{0.1}$ is found from a fit to our computed models. The parameter $\alpha$ controls the numerical diffusion of the electric field, if we assume that numerical diffusion mimics the physical one to some extent \citep[i.e. $\alpha\propto 1/\eta$, where $\eta$ is the resistivity, see discussion Sect.~\ref{sec:discussion} and][]{Mahlmann2020c}. Thus, our results suggest that the luminosity of the aligned rotator is inversely proportional to the resistivity at the ECS. Nevertheless, because of the small powerlaw index in the aforementioned $\alpha$-luminosity-relation, large changes in the resistivity are required to produce significant variations of the luminosity at the LC.

\subsection{Energy dissipation beyond the Y-point}
\label{sec:econservation}

\begin{figure}
  \centering
  \includegraphics[width=0.47\textwidth]{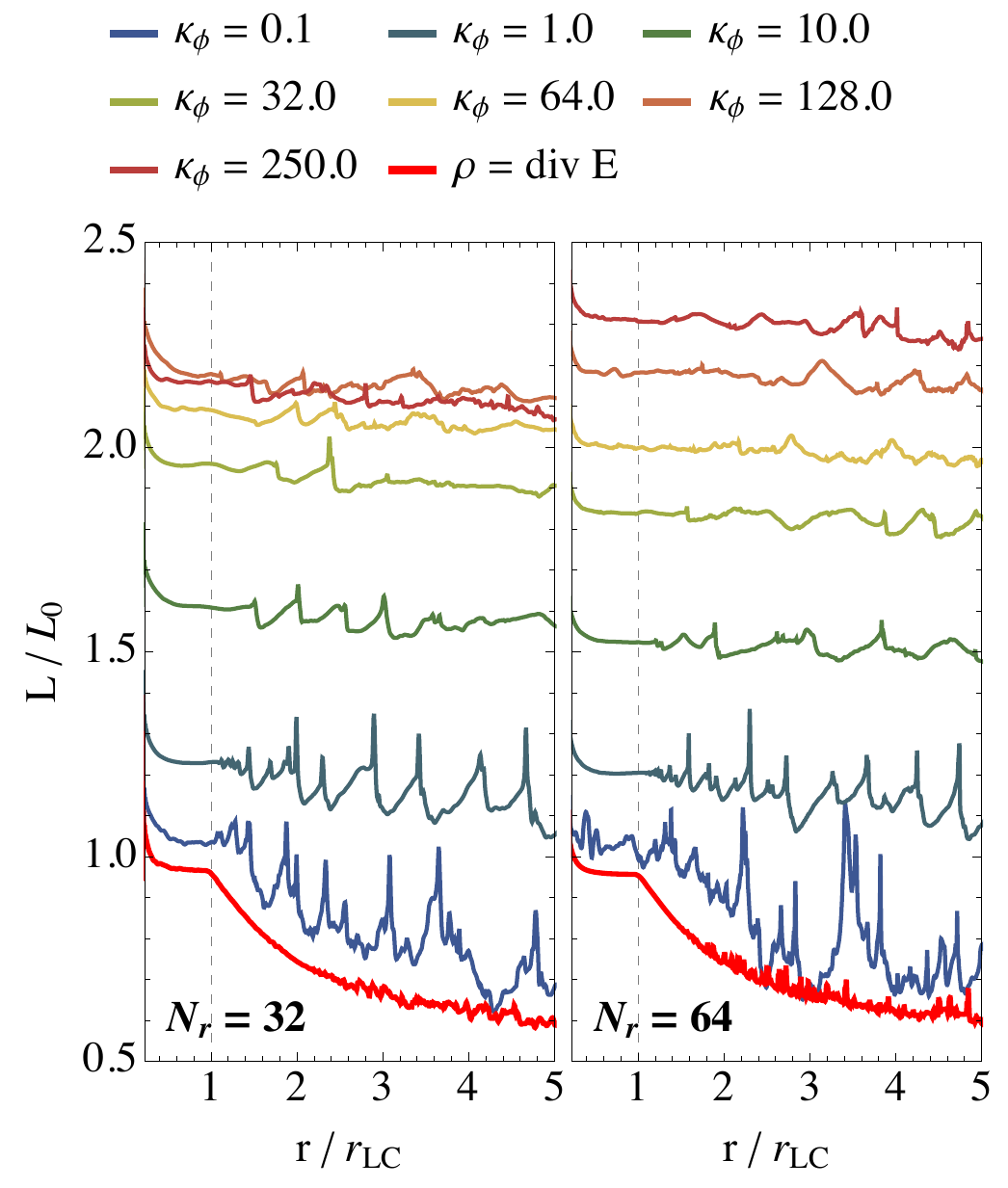}
  \vspace{-6pt}
  \caption{Poynting flux as a function of the distance from the central rotator. We present results after $16.7$ rotation periods for different damping constants $\kappa_\Phi$ (and a constant advection parameter $c_\Phi=1$) for different resolutions. The tests employing the LCR method are indicated by thick red lines.}
\label{fig:Poynting}
\end{figure}

\citet{Tchekhovskoy2013} hold a reliable benchmark for the convergence of pulsar magnetosphere modeling and its comparison between FFE and MHD. We emphasise the following property of their results: MHD models show a decay of the Poynting flux beyond the light cylinder that diminishes with higher resolution. However, their FFE models of the aligned rotator magnetosphere show $43\% - 50\%$ dissipation that converge to a stable amount of dissipation ($43\%$) for the highest resolution. We observe such behavior for the case of LCR, as it is marked by the thick red lines in Fig.~\ref{fig:Poynting}. 

In all other cases using the CC method, the dissipation of Poynting flux along the radial direction is significantly less. This observation is also supported by
the field line images of Fig.~\ref{fig:FLDSJDENS}. There, field lines notably reconnect beyond the LC in the limit of low $\alpha$, while the cleaning of numerical errors shapes the ECS over longer distances. Stationary solutions of the force-free aligned rotator magnetosphere involving an ECS (where the FFE approximation does not hold) have been obtained \citep[e.g.][]{Contopoulos_2007A&A...466..301, Contopoulos_2014ApJ...781...46, Contopoulos2020}. The configurations built by \citet{Contopoulos2020} show that there is a region of the ECS that may extend beyond the LC, where magnetic field lines can still be closed. Our time-dependent models certainly reveal such a region, and its properties depend on the model parameters, which ultimately determine the level of (numerical) dissipation of the method. Small-scale structures in the equatorial plane (such as plasmoid-like formations) do not automatically appear by increasing resolution. Rather, it seems that an efficient damping of charge conservation errors allows them to emerge.

Remarkably, not only the smallest luminosity $L_{\rm LC}$ is attained by the LCR method, but, most importantly, also the largest relative decrease of the luminosity, $\Delta L/L_{\rm LC} = \left[L(r=5r_{\rm LC})- L_{\rm LC}\right]/L_{\rm LC}$ between the light cylinder and $r=5r_{\rm LC}$. Taking $\Delta L/L_{\rm LC}$ as a measure of the diffusivity of the algorithm, we conclude that the LCR method is more diffusive than the CC method. This conclusion is numerically robust, since duplicating the spatial resolution does not notably change the dissipation beyond the Y-point observed in the individual models.

\begin{figure}
  \centering
  \includegraphics[width=0.47\textwidth]{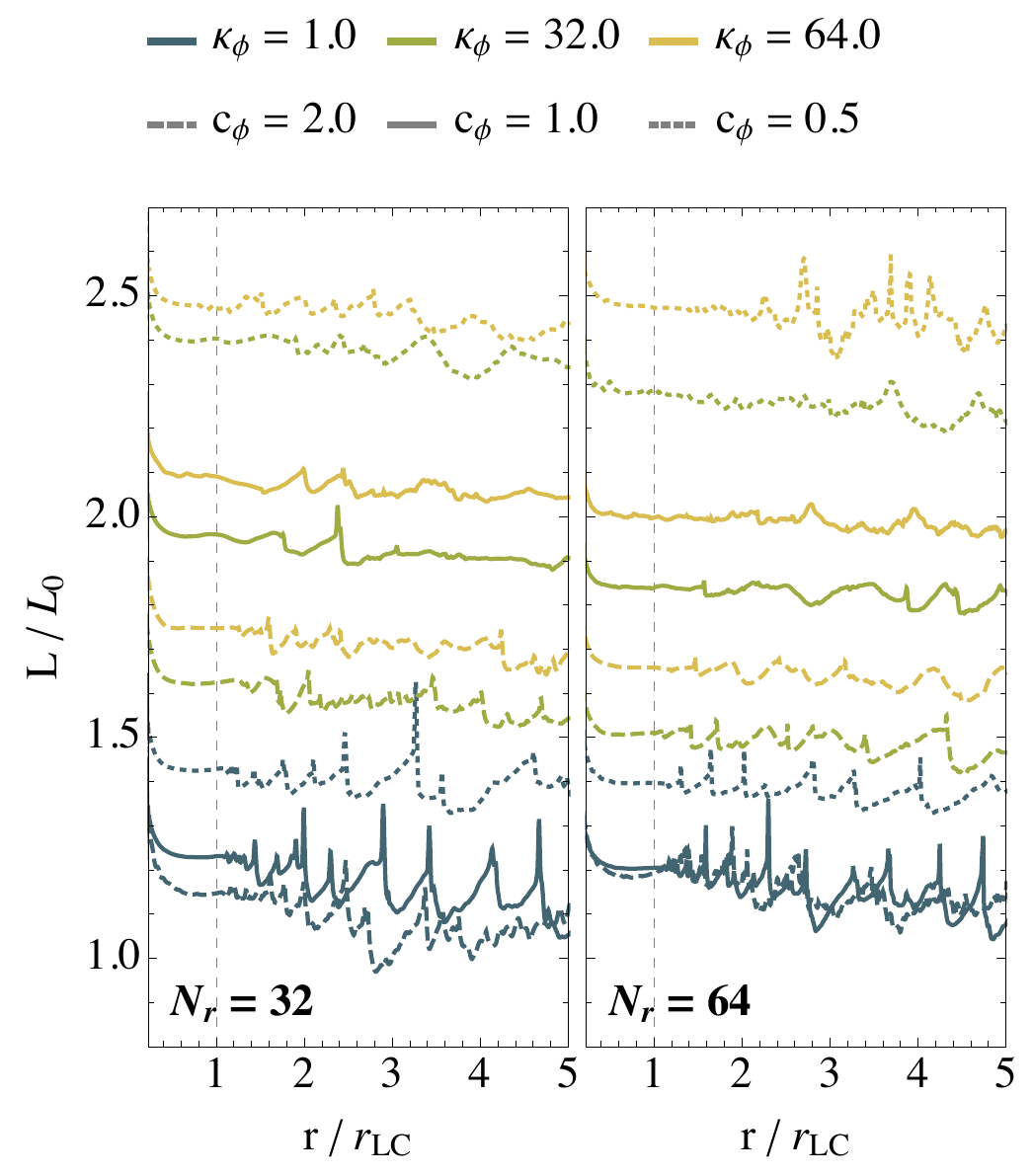}
  \vspace{-6pt}
  \caption{As Fig.~\ref{fig:FLDSJDENS} but for selected damping constants $\kappa_\Phi$ and different choices for the advection parameter $c_\Phi$.}
\label{fig:PoyntingCSPEED}
\end{figure}

Figs.~\ref{fig:Poynting} and~\ref{fig:PoyntingCSPEED} show a significant variability of the Poynting flux beyond the light cylinder for small values of $\kappa_\Phi$. 
For most values of $\kappa_\Phi$, the Poynting flux beyond the LC is not monotonically decaying, but shows spikes associated to the ejection of plasmoid-like structures that move \emph{outwards} along the ECS (see Fig.~\ref{fig:FLDSJDENS}). The fact that these blobs of strong currents move outwards even if they are produced inside the LC is relevant: they do not contribute to the growth of the closed magnetospheric region. This finding is in contrast to \citet{Komissarov_2006MNRAS.367...19}, who claims that part of the plasmoids will move inwards, increasing the size of the closed magnetosphere with time and, hence, pushing the Y-point towards the LC. This assertion has also led \cite{Spitkovsky2005} and \cite{McKinney2005} to suggest that all configurations with $x_0<1$ would be unstable or transitory \citep[see also][]{Kalapotharakos_2009A&A...496..495,Yu2011,Parfrey_2012MNRAS.423.1416}. We do not notice such behavior here and, thus, our results suggest that closed magnetospheric configurations with a Y-point inside the LC may survive for many dynamical times.

In order to compute $\Delta L/L_{\rm LC}$ when it is spatially (and temporarily) variable (i.e., using the CC method with small values of $\alpha$), it is necessary to smooth out the data taking a suitable moving average. Larger values of $\kappa_\Phi$ or smaller advection speeds $c_\Phi$ (i.e., larger values of $\alpha$) yield magnetospheres with smaller Poynting flux dissipation beyond the LC. We note that the dissipation takes place at the ECS, where the force-free approximation is not strictly valid, specifically because the electric field strength becomes larger than the magnetic one and non-ideal electric fields with $\mathbf{E}\cdot\mathbf{B}\neq 0$ become dynamically important. Hence, the different amounts of relative dissipation are closely connected with the numerical handling of the force-free constraints and the propagation of errors from the regions where FFE is breached.

Extreme values of $\alpha$ suppress the hyperbolic/parabolic cleaning, letting the charge and the corresponding electric field divergence become misaligned over time. Nevertheless, very strong damping cleans errors very rapidly but effectively decouples the evolution equations of $\Phi$ from the underlying system of physical balance laws (in this case, Maxwell's equations), and renders the scalar potential dynamically negligible. As cases of intermediate $\alpha$ do not show this variability, we empirically demonstrate that there is an optimal range of values of $\kappa_\Phi$, corresponding to $\alpha\sim 1$. In this range, the CC method works very efficiently, effectively minimizing the dissipation at the ECS. We find strong evidence for a much better conservation of energy flux beyond the LC than what is quoted throughout the literature, where results seem to correspond to the most diffusive case of our parameter exploration, as to say the LRC method.
 As a matter of fact, for the intermediate values of $\alpha\in\left[1.45, 4.63\right]$, the current $\mathbf{j}_\parallel$ is very efficiently suppressed in the equatorial region beyond the Y-point (Fig.~\ref{fig:FLDSJDENS}). Indeed, this is the reason which explains the significantly reduced dissipation beyond the LC in the models using the CC method and optimal values of $\alpha\sim 1$. For the smallest values of $\alpha$, the variability timescales are of the order of the polar cap light-crossing time, namely $\sim r_*\theta_{\rm c}/c\sim r_*/2c$ (see also the discussion in Sect.~\ref{sec:diffusivitydiscuss}).

\begin{figure*}
  \centering
  \includegraphics[width=0.85\textwidth]{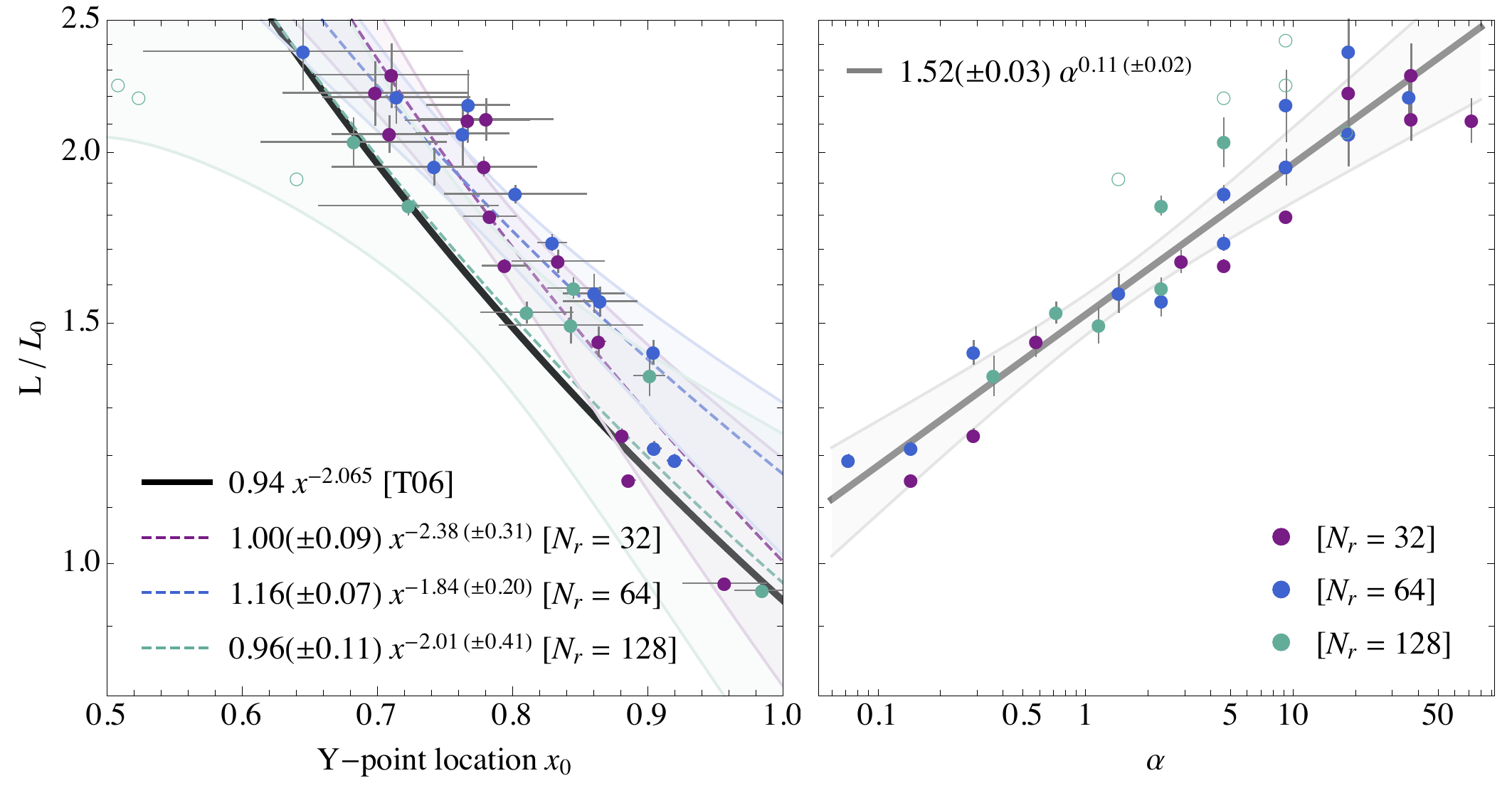}
  \vspace{-6pt}
  \caption{Dependence of the luminosity on the Y-point location and diffusion parameter $\alpha$. For the Y-point location (left panel), measurements from simulations at five different moments in time are averaged (standard deviation indicated in error bars). Fit functions are derived from for all models that find a reliable equilibrium (solid dots) and are represented by colored dashed lines. As a reference, we compare these results to the corresponding relation in \citet[][black line]{Timokhin2006}.
  Errors to these fit functions are represented by the lightly shaded regions of the respective color. Outliers that do not find stable equilibria are denoted by colored circumferences. The right panel shows the normalised luminosity measured at the light cylinder, evaluated against the diffusion parameter $\alpha$. A fit function to the data is shown with a thick solid line. Uncertainties to the fit function are given as a shaded region around the central line.}
\label{fig:YpointLuminosity}
\end{figure*}

\section{Sources of dissipation}
\label{sec:tradesecrets}

In this part of the manuscript, we present an array of ancillary high-resolution simulations that will be key for assessing the role of numerical and/or phenomenological diffusivity in shaping the overall magnetospheric structure, and disclosing several \emph{trade secrets} \citep[cf.][]{Contopoulos2016} in the dynamical modeling of aligned rotator magnetospheres. In order to speed up the calculations at the higher numerical resolution, (i.e. with a grid spacing $\Delta r=r_*/128$, $\Delta\theta=\pi/400$) we will employ a radially re-scaled coordinate system for the remainder of this section. Specifically, for $r\gtrsim 3r_{\rm LC}$, the grid spacing increases by a factor of $a=1.001$ in each grid point along the radial direction. This helps us to reduce the number of points enclosed by the simulation domain while keeping sufficient resolution around the central object. Besides this change introduced for numerical convenience, we further replace the boundary layer described in Sect.~\ref{sec:simulationsetup} by perfect conductor boundary condition within the Riemann solver \citep[cf.][]{Munz2000}. Specifically, at the inter-cell face where the stellar surface is located, we set the following 'left' (L) state of the Riemann problem (corresponding to the interior of the star):
\begin{align}
\Psi_L &= \Psi_R\\
\Phi_L &= -\Phi_R\\
\mathbf{E}_L &= -\mathbf{E}_R + 2(\mathbf{E}_R \cdot \mathbf{\hat r}) \mathbf{\hat r}\\ 
\mathbf{B}_L &= +\mathbf{B}_R - 2(\mathbf{B}_R \cdot \mathbf{\hat r}) \mathbf{\hat r} ,
    \label{eq:BCs}
\end{align}
where $\mathbf{\hat r}$ is the unit radial vector (normal to the stellar surface) and R denotes the respective 'right' state. The reason to consider different boundary conditions is the fact that we seek 
stationary (or nearly stationary) magnetospheric configurations, in which case boundary conditions nearly completely determine the structure of the magnetosphere. Hence, using a variety of boundary conditions allows as to gauge their influence on the most salient features of our solutions, namely, the location of the Y-point and the closely related dissipation at the ECS. As we shall see, the boundary strategies used throughout this work produce consistent results. Table~\ref{tab:ancilmodels} shows a list of all the simulations that we describe in the following subsections.

\begin{table}
\centering
\caption{Properties of the set of high-resolution simulations corresponding to models described in Sect.~\ref{sec:tradesecrets}. We provide the label of the respective model, the strategies used to deal with violations of the force-free conditions, the strategies to model the electric charge, the phenomenological resistivity induced by a suitable current, the Y-point location, the luminosity at the LC, its relative decay up to a distance of $5r_{\rm LC}$, and the colatitude of the closed zone separatrix at the stellar surface. Models $\mathbf{Bg}$ to $\mathbf{Br}$ employ $\kappa_I=8$ in the parametrization of Eq.~(\ref{eq:FFResCurrentPerpendicular}).} 
\label{tab:ancilmodels}
\begin{tabular}{P{0.2cm}P{1.2cm}P{0.4cm}P{0.5cm}P{0.5cm}P{0.7cm}P{0.8cm}P{0.4cm}}
\hline
\hline
 & $\mathbf{E}\cdot\mathbf{B}$ & $\rho$ & $\alpha$ & $x_0$ & $L_\Yp/L_0$ & $\Delta L/L_\Yp$ & $\theta_{\rm pc}$\\
\hline
\textbf{Ba} & algebraic & LCR & $4.6$ & $1.04$ & $0.96$ & $0.29$ & $31.5^\circ$\\%$0.55$ \\
\textbf{Bb} &  algebraic  & CC & $4.6$ & $0.69$ & $2.10$ & $0.03$ & $37.2^\circ$\\%$0.65$ \\
\textbf{Bc} &  algebraic  & HCC & $4.6$ & $0.68$ & $2.04$ &$<0.01$ & $36.7^\circ$\\%$0.64$ \\
\textbf{Bd} & algebraic & LCR & $0.7$ & $1.02$ & $0.96$ & $0.32$ & $31.5^\circ$\\%$0.55$ \\
\textbf{Be} & algebraic &  CC & $0.7$ & $0.83$ & $1.52$ & $0.02$ & $34.4^\circ$\\%$0.60$ \\
\textbf{Bf} & algebraic & HCC & $0.7$ & $0.74$ & $1.49$ & $<0.01$ & $35.0^\circ$\\%$0.61$ \\
\hline
\textbf{Bg} & $\eta = 0.0$ &  LCR & $4.6$ & $1.04$ & $0.95$ &  $0.31$ & $31.5^\circ$\\%$0.55$ \\
\textbf{Bh} & $\eta = 0.0$  & HCC & $4.6$ & $0.81$ & $1.68$ & $<0.01$ & $34.4^\circ$\\%$0.60$ \\
\textbf{Bi} & $\eta = 0.0$ &  LCR & $0.7$ & $1.04$ & $0.95$ & $0.30$ & $31.5^\circ$\\%$0.55$ \\
\textbf{Bj} & $\eta = 0.0$ & HCC & $0.7$ & $0.94$ & $1.11$ & $<0.01$ & $31.5^\circ$\\%$0.55$ \\
\hline
\textbf{Bk} & $\eta = 1.0$ & HCC & $0.7$ & $0.91$ & $1.15$ & $0.02$ & $31.5^\circ$\\%$0.55$ \\
\textbf{Bl} & $\eta = 10^{-1}$ & HCC & $0.7$ & $0.93$ & $1.10$ & $0.02$ & $31.5^\circ$\\%$0.55$ \\
\textbf{Bm} & $\eta = 10^{-2}$ & HCC & $0.7$ & $0.95$ & $1.10$ & $0.01$ & $30.9^\circ$\\%$0.54$ \\
\textbf{Bn} & $\eta = 10^{-3}$ & HCC & $0.7$ & $0.97$ & $1.10$ & $0.01$ & $30.9^\circ$\\%$0.54$ \\
\textbf{Bo} & $\eta = 1.0$ &  CC & $0.7$ & $0.90$ & $1.30$ & $<0.01$ & $32.7^\circ$\\%$0.57$ \\
\textbf{Bp} & $\eta = 10^{-1}$ &  CC & $0.7$ & $0.91$ & $1.29$ & $<0.01$ & $32.7^\circ$\\%$0.57$ \\
\textbf{Bq} & $\eta = 10^{-2}$ &  CC & $0.7$ & $0.91$ & $1.28$ & $0.01$ & $32.7^\circ$\\%$0.57$ \\
\textbf{Br} & $\eta = 10^{-3}$ &  CC & $0.7$ & $0.90$ & $1.29$ & $0.02$ & $32.7^\circ$\\%$0.57$ \\
\hline
\textbf{Bs} & algebraic & LCR & $0.0$ & $0.47$ & $1.34$ & $0.42$ & $31.5^\circ$\\%$0.55$ \\
\textbf{Bt} & algebraic & LCR & $0.1$ & $0.52$ & $1.03$ & $0.32$ & $31.5^\circ$\\%$0.55$ \\
\textbf{Bu} & algebraic & LCR & $2.3$ & $1.02$ & $0.96$ & $0.32$ & $31.5^\circ$\\%$0.55$ \\
\textbf{Bv} & algebraic & LCR & $9.3$ & $1.02$ & $0.96$ & $0.32$ & $31.5^\circ$\\%$0.55$ \\
\textbf{Bw} & algebraic & LCR & $18$ & $1.02$ & $0.96$ & $0.31$ & $31.5^\circ$\\%$0.55$ \\
\hline
\hline
\end{tabular}
\end{table}

\subsection{The role of the violations of FFE constraints}
\label{sec:focuseddominance}

\begin{figure*}
  \centering
  \includegraphics[width=0.98\textwidth]{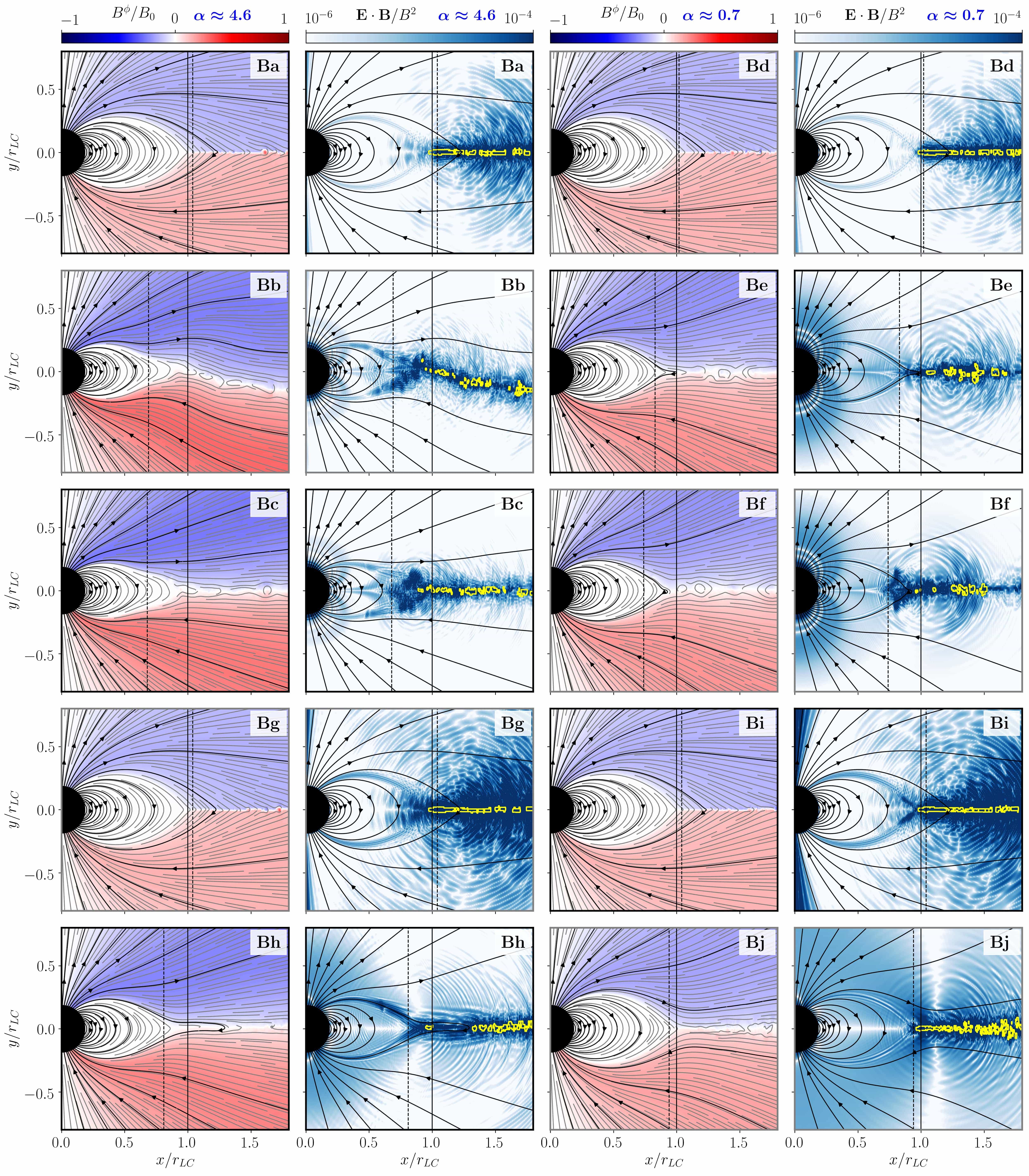}
  \caption{Comparison of different treatments of the charge $\rho$ for some
   configurations specified in Tab.~\ref{tab:ancilmodels}. We display the toroidal magnetic field in the left panel of each respective model. The corresponding right panels visualise the force-free violations accumulating during one full time-step of the FFE integration. We show the growth of non-ideal electric fields (absolute value) emerging transiently due to the violation of the $\mathbf{E}\cdot\mathbf{B}=0$ constraint by blue colors, and locations where the magnetic dominance condition is breached by yellow contours. The vertical solid and dashed black lines denote the position of the LC and of the Y-point, respectively.
  }
\label{fig:FFCOND_COMPARE}
\end{figure*}

\begin{figure}
  \centering
  \includegraphics[width=0.47\textwidth]{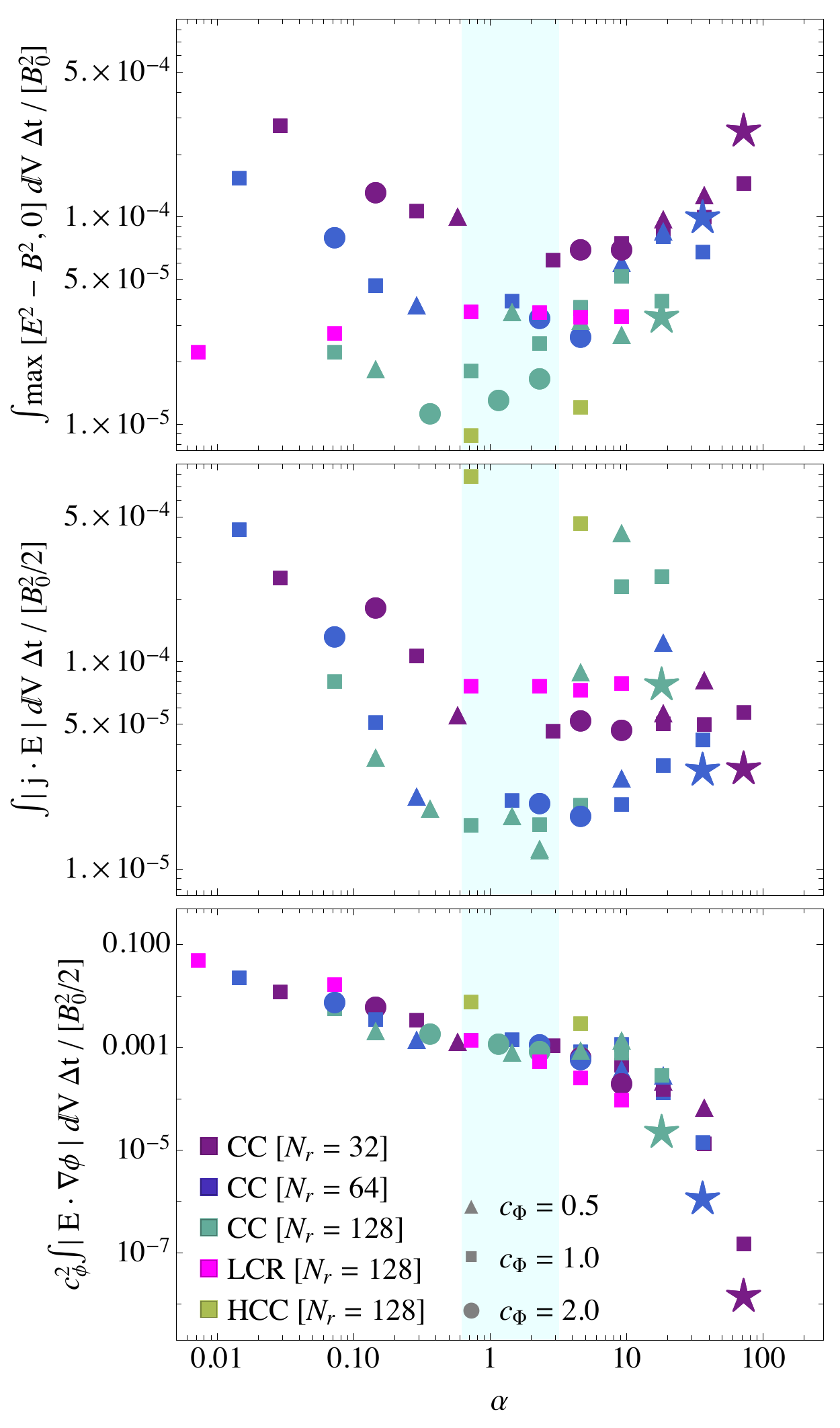}
  \vspace{-6pt}
  \caption{Dissipation via different channels of numerical diffusion as a function of the dimensionless parameter $\alpha$ (Eq.~\ref{eq:alphadimensionless}) in a quasi-equilibrium state at $t\approx 7.5t_p$. An estimate of the favorable range for $\alpha$ that minimises the dissipation by force-free violations is indicated as a light blue region. Different colors of the markers indicate different resolutions and treatment of charge density evolution. The shapes of the markers vary with different $c_\Phi$. We provide reference values of the respective dissipation channel for the LCR method of highest $\alpha$ as star-shaped data-points.}
\label{fig:QDissipation}
\end{figure}

We compare the final (quasi-)equilibrium of three different numerical treatments of the electric charge in the magnetosphere in models $(\mathbf{Ba})-(\mathbf{Bf})$. Since we have introduced changes in the boundary conditions and in the numerical grid, we confirm the results from the previous section (incorporating the aforementioned modifications with respect to models in Sect.~\ref{sec:forcefreealigned}). First, we use the LCR method throughout the entire magnetosphere for two distinct values of $\alpha$ (models $\mathbf{Ba}$ and $\mathbf{Bd}$). Second, we use the CC method in models $\mathbf{Bb}$ and $\mathbf{Be}$.
Finally, we apply the HCC method in models $\mathbf{Bc}$ and $\mathbf{Bf}$ using different values of $\alpha$.

Figure~\ref{fig:FFCOND_COMPARE} demonstrates that the cases where charge is not conservatively evolved in the entire domain (models $\mathbf{Ba}$ and $\mathbf{Bd}$) produce congruent results independent of the cleaning parameter $\alpha$. Such solutions have a large amount of reconnecting field lines beyond the light cylinder and a luminosity at $r_{\rm LC}$ that approaches $L_\LC/L_0\sim 1$. The small relative differences observed in the position of the Y-point ($\lesssim 2\%$) and the insignificant change in $L_{\rm LC}/L_0$ compared with the $\simeq 10\%$ relative difference in $\Delta L/L_{\rm LC}$ highlight the fact that the variation of $\alpha$ (by a factor $\approx 6.6$) mostly affects the dissipation dynamics of the ECS and its neighborhood. 
When charge is evolved by a separate continuity equation, any misalignment of the electric field divergence and the charge density is cleaned out by the scalar potential $\Phi$. As we presented above (Sect.~\ref{sec:forcefreealigned}), altering the cleaning parameter $\alpha$ shifts the position of the Y-point and changes the Poynting flux at $r_{\rm LC}$. While most of the cases employing the CC method have a well-maintained current sheet beyond the Y-point (contrasting the reconnection in models $\mathbf{Ba}$ and $\mathbf{Bd}$), excessive cleaning (driven by large values of $\alpha$) can induce variability of the current sheet along the vertical direction. 
This is obvious when comparing models $\mathbf{Bb}$ and $\mathbf{Be}$, but can also be observed for large values of $\alpha$ in Fig.~\ref{fig:FLDSJDENS}. As we see when comparing models $\mathbf{Bb}$ and $\mathbf{Bc}$, using the HCC scheme stabilises the current sheet on the equator, damping vertical displacements in the ECS. Inside the LC, models $\mathbf{Bc}$ and $\mathbf{Bf}$ only slightly differ from their CC counterparts. The location of the Y-point, and the luminosity differ less than $\simeq 2\%$ among HCC and CC models for $\alpha=0.7$, while for $\alpha=4.6$, HCC models show values of $x_0$ and $L_\LC/L_0$ in between of the ones of CC and LCR models (Tab.~\ref{tab:ancilmodels}). Beyond the LC, the differences between HCC and CC models are driven by insufficient damping, especially in the model with larger $\alpha$ ($\mathbf{Bb}$), where the ECS is distorted and the violations of the magnetic dominance condition are more patchy and intermittent along it. The fact that HCC models $\mathbf{Bc}/\mathbf{Bf}$ look more like $\mathbf{Bb}/\mathbf{Be}$ than like $\mathbf{Ba}/\mathbf{Bd}$, respectively, results from a subtle combination of two effects. On the one hand, the corrections to the electric fields after violations of the magnetic dominance condition are larger when using a local reconstruction of charge, either globally (LCR) or locally (HCC), than in CC models. This seems natural as the largest charges are created as a result of the sharp discontinuities of the electric field across the ECS, where LCR and HCC models undergo the same corrections in the charge density. On the other hand, the commonality in the (hyperbolic) evacuation of the errors triggered by violations of the FFE constraints at the ECS results in a greater similarity between HCC and CC models than to LCR models (the feedback of $\nabla\cdot\mathbf{E}$ on $\Psi$ is significantly reduced because of the explicit form in which $\rho$ is constrained in the LCR method; Sect.~\ref{sec:numericalmethod}).  We can, thus, conclude that the large diffusivity in case of LCR method is primarily induced by corrections to the electric fields after violations of the magnetic dominance condition.
 
The (algebraic) correction of violations to the force-free conditions has different consequences for each of the constraints. Deviations from the $\mathbf{E}\cdot\mathbf{B}=0$ condition build up continuously and are distributed throughout the domain, though they are larger close to current sheets. We visualise this in the panels of Fig.~\ref{fig:FFCOND_COMPARE} that show the  $\mathbf{E}\cdot\mathbf{B}$ errors normalised to the local magnetic field strength in a blue gradient. Violations include both very small inaccuracies that result from truncation errors of the algorithm, and strong non-ideal electric fields emerging, for example, at current sheets. In models employing the LCR method ($\mathbf{Ba}$ and $\mathbf{Bd}$), the hyperbolic part of the cleaning still operates to drive violations of the  $\mathbf{E}\cdot\mathbf{B}$ constraint away from the ECS (we note a \emph{wave} pattern - concentric structures - apparently emerging from the equatorial plane at about $r\approx 1.2 r_{\rm LC}$ in the respective panels of Fig.~\ref{fig:FFCOND_COMPARE}). That pattern persists in models $\mathbf{Bb}$ and  $\mathbf{Be}$, as well as in $\mathbf{Bc}$ and  $\mathbf{Bf}$, where we employed the CC and the HCC method. However, in these cases, the violations of  $\mathbf{E}\cdot\mathbf{B}$ can affect a larger region, especially for small values of $\alpha$. The mid-panel of Fig.~\ref{fig:QDissipation} shows that models employing the CC method systematically dissipate less energy by Ohmic processes (here approximated by the amount of current parallel to the electric field added up over the whole domain on a single timestep) than models using the LCR method for $\alpha\lesssim 5$.

The condition $\mathbf{E}^2-\mathbf{B}^2\leq 0$ is only relevant when significant non-ideal electric fields have built up, in the setup at hand at the ECS, as it is lucidly illustrated by the yellow contours in Fig.~\ref{fig:FFCOND_COMPARE}. It is a natural impulse to associate the high diffusivity observed for the setups using the LCR method to the violations of the magnetic dominance. Looking at the time evolution of the violation of the magnetic dominance constraint, we observe that in models implementing the LCR method everywhere, one finds regions of $\mathbf{E}^2>\mathbf{B}^2$ that consistently cover the ECS in the range $x_0\lesssim r/r_{\rm LC}\lesssim 1.7$ (and likely beyond). In contrast, models employing the CC and HCC methods display an intermittent set of spots of smaller extension, where $\mathbf{E}^2>\mathbf{B}^2$ in the equatorial region beyond the Y-point. The top panel of Fig.~\ref{fig:QDissipation} displays the electric energy subtracted from the whole domain during one iteration of the time-integrator. Contributions to this channel of diffusion result from mesh cells where the magnetic dominance constraint is breached. In practice, we calculate averages from several snapshots of the results to ensure that the quoted dissipation estimates have (more or less) stabilised. LCR models computed at the highest resolution ($N_r=128$; magenta squares) systematically dissipate more (the least about the same) energy by restoring the magnetic dominance condition than models computed with the CC method and the same resolution (dark green symbols) for $\alpha\lesssim 3$. Similarly to the electric energy dissipation in the correction of the  $\mathbf{E}\cdot\mathbf{B}=0$ constraint, there is an interval $0.5 \lesssim \alpha\lesssim 3$, where dissipative losses induced by the restoration of the magnetic dominance condition are minimised. 

In contrast to the trend found for the CC method, the dissipation triggered by violations of the FFE constraints is quite insensitive to $\alpha$ when evaluated for the LCC method (see magenta symbols in the upper and mid panels of Fig.~\ref{fig:QDissipation}). In order to understand this behavior, we have computed the electric energy dissipated by the hyperbolic/parabolic cleaning algorithm, which is given by $|c_\Phi^2\mathbf{E}\cdot\nabla \Phi|$ in the lower panel of Fig.~\ref{fig:QDissipation}. While the gradients of the cleaning potential $\Phi$ are large enough, increasing $\alpha$ (i.e. damping the divergence errors faster than shifting them away) reduces the diffusion through this channel, independent of the method used to evolve or reconstruct the charge density (LCR or CC). The dissipation through this channel is significantly larger than that driven by violations of the FFE constraints for $\alpha\lesssim 5$ and dominates the overall diffusion of electromagnetic energy in the magnetosphere. Above that value of $\alpha$, the total dissipated energy in the $|c_\Phi^2\mathbf{E}\cdot\nabla \Phi|$ channel is smaller than in the other two channels (for this, we compare the numerical values in the top and mid panels of Fig.~\ref{fig:QDissipation} to those in the bottom panel). From that point on, the cleaning algorithm in the CC method cannot efficiently evacuate and damp the errors induced in $\nabla\cdot\mathbf{E}$ by the restoration of the FFE constraints. Thus, the energy dissipation in models employing the CC method, becomes dominated by the violation of the FFE constraints above a certain value of $\alpha$. This reasoning substantiates our claim about the existence of an optimal interval for $0.7\lesssim\alpha\lesssim 3$, where the overall dissipation in the magnetosphere is smaller using the CC method than the LCR one. The insensitivity of the dissipation due to violations of the FFE constrains on $\alpha$ for LCR models (for a fixed numerical resolution) can be explained by the limited evacuation of FFE violations, e.g., from current sheets. As $\rho\approx\nabla\cdot\mathbf{E}$ (up to discretization errors), the right-hand-side of Eq.~\eqref{eq:Phi} is significantly smaller than in the models equipped with the CC method; the two mechanisms (cleaning of divergence errors and enforcing FFE violations) operate independently.

\subsubsection{Algebraic corrections versus driver currents}
\label{sec:drivingfocus}

\begin{figure*}
  \centering
  \includegraphics[width=1.0\textwidth]{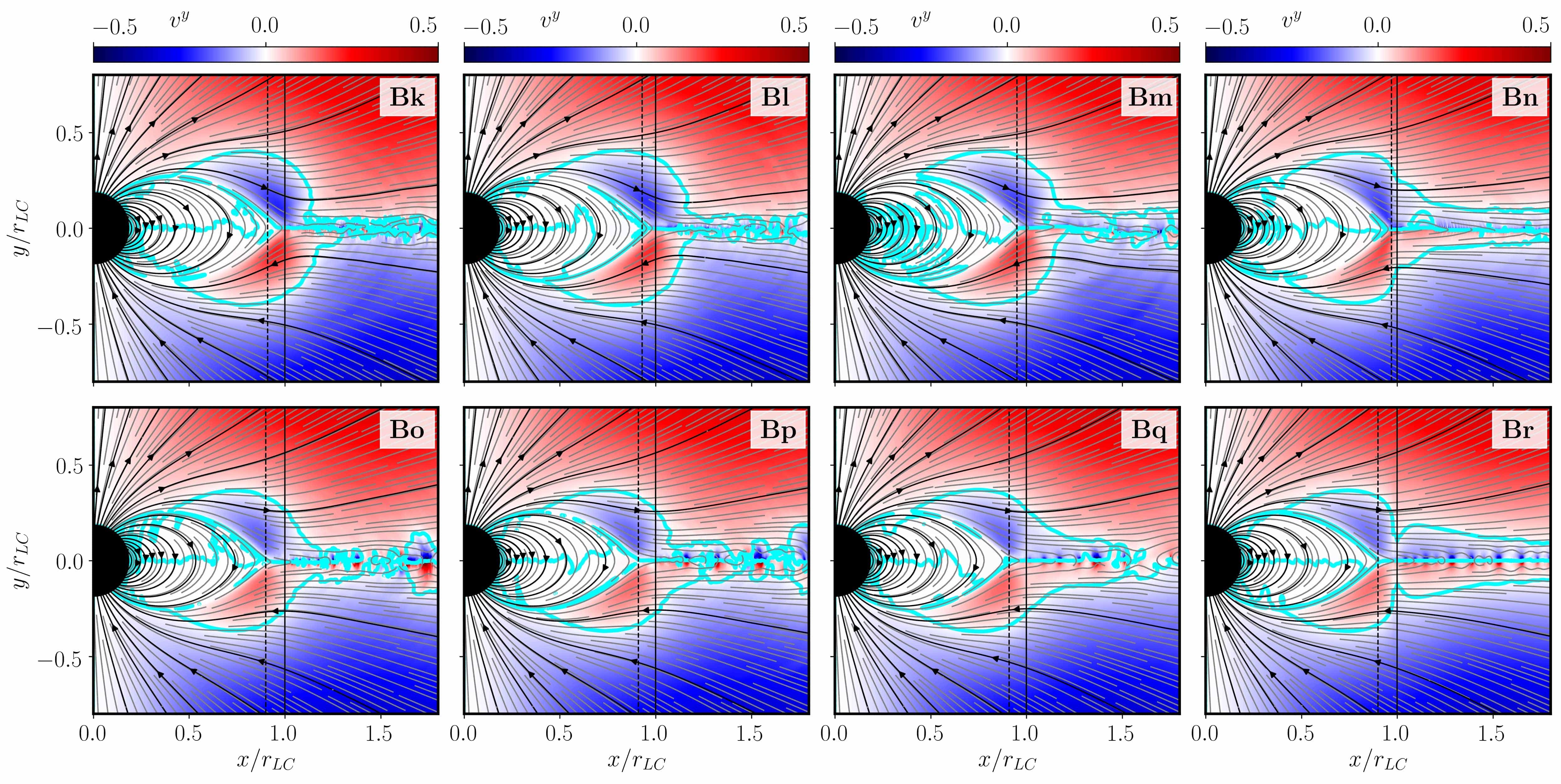}
  \vspace{-18pt}
  \caption{Comparison of magnetospheric dynamics for different values of $\eta$ in the current given by Eq.~(\ref{eq:FFResCurrentPerpendicular}).  From \textit{left} to \textit{right}, the phenomenological resistivity is increasing as $\eta\in\left\{10^{-3},10^{-2},10^{-1},1.0\right\}$. The background color denotes the (drift) velocity component $v^y$, as to say the velocity pointing perpendicular to the ECS. The separatrices where $v^y=0$ are highlighted by magenta contours. The top row comprises models implementing the HCC method, the bottom row those of fully conservative evolution of the charge density $\rho$ (CC models).}
\label{fig:ETA_COMPARE}
\end{figure*}

We have seen in the previous section that violations of the FFE constraints are fundamentally connected to the magnetospheric structure. It is, therefore, natural to assess whether different methods to restore the FFE constraints impact on the overall dissipation and topology of the magnetosphere.
Throughout the literature, various techniques are used to correct any deviations from the $\mathbf{E}\cdot\mathbf{B}=0$ condition. These techniques split up into the ones that apply algebraic resets of the electric field at each violation (as done in the models shown so far), and those that modify the Ohm's law encoded in a suitable current to drive the system into a force-free state \citep[cf.][and references therein]{Mahlmann2020b}. It is very much justified to suspect that all the variability in the luminosity identified in Sect.~\ref{sec:forcefreealigned} stems from this critical ingredient to force-free codes, rather than from their connection to charge conservation. 
To dissect this subtle issue, we conduct simulations of the models $(\mathbf{Bg})-(\mathbf{Bh})$. Without algebraic resets of the $\mathbf{E}\cdot\mathbf{B}=0$ condition, we employ $\kappa_I=8$ in the force-free current presented in Eq.~(\ref{eq:FFResCurrentPerpendicular}). In this way, our method is comparable to the ones that employ driving currents, namely, a formulation of Ohm's law with a finite resistivity $\sigma_\parallel$ that acts along the direction of the magnetic field \citep[e.g.,][]{Alic2012,Komissarov2004}. 

In Fig.~\ref{fig:FFCOND_COMPARE}, we contrast the magnetospheric equilibrium found when combining a suitable current to drive $\mathbf{E}\cdot\mathbf{B}\rightarrow 0$ for the LCR method (models $\mathbf{Bg}$ and $\mathbf{Bi}$), and for the HCC method (models $\mathbf{Bh}$ and $\mathbf{Bj}$). One straightforward observation is that a local reconstruction of charge remains the most diffusive limit in view of the much larger decrease of luminosity with distance measured by $\Delta L/L_{\rm LC}$ (Tab.~\ref{tab:ancilmodels}). For the HCC models $\mathbf{Bh}$ and $\mathbf{Bj}$, the luminosity decrease beyond the LC is as small as in the case in which algebraic cutbacks of the electric field enforce $\mathbf{E}\cdot\mathbf{B}=0$ (models $\mathbf{Bc}$ and $\mathbf{Bf}$). The luminosity of the HCC models that implement a driving current to enforce $\mathbf{E}\cdot\mathbf{B}\rightarrow 0$ decreases by $\sim 18\%-25\%$ with respect to models using algebraic resets of the electric field (models $\mathbf{c}$ and $\mathbf{f}$). This is a direct consequence of the larger value of $x_0$ in models $\mathbf{Bh}$ and $\mathbf{Bj}$, which place their Y-points closer to the LC. In contrast, models using the LCR method display only a small change of the luminosity ($\lesssim 1\%$) when comparing models with different strategies to enforce $\mathbf{E}\cdot\mathbf{B}=0$ (i.e. models $\mathbf{Ba}/\mathbf{Bd}$ vs models $\mathbf{Bg}/\mathbf{Bi}$). Thus, the smaller luminosity of models using the LCR method is not fully accounted for by the algebraic - rather harsh - corrections we apply to non-ideal electric fields.

\subsubsection{Diffusivity models beyond the light cylinder}
\label{sec:diffusivityfocus}

One technique that is often associated to the capacity of an FFE scheme to resolve current sheets is a finite resistivity induced by a suitably chosen Ohm's law \citep{Alic2012,Parfrey2017}. In \citet{Mahlmann2020c}, we explore the action of the current prescribed by Eq.~(\ref{eq:FFResCurrentPerpendicular}) during the development of 2D tearing modes. In this section, we review the impact of such phenomenological resistivities on the current sheet of the aligned rotator magnetosphere. To this purpose, we prescribe the following phenomenological resistivity:
%parameter
%
\begin{align}
    \eta=\eta_{\rm bg}+(\eta_{\rm d}-\eta_{\rm bg})\, \frac{1+\tanh(r_{\rm cyl}-r_{\rm LC})}{2} .
\end{align}
This resistivity becomes the driving value $\eta_{\rm d}$ for $r_{\rm cyl}>r_{\rm LC}$, where $r_{\rm cyl}=r\sin\theta$ is the cylindrical radius. In other words, inside the LC, all tests have the same (small) background resistivity $\eta_{\rm bg}$, where we use $\eta_{\rm bg}=10^{-5}\ll \eta_{\rm d}$.

In Fig.~\ref{fig:ETA_COMPARE} 
%and~\ref{fig:ETA_COMPARE_Q10} 
we compare the magnetospheric evolution of different values of $\eta_{\rm d}$ for the HCC and CC models (\textbf{Bk})-(\textbf{Bn}) and (\textbf{Bo})-(\textbf{Br}), respectively. As we also denote in Tab.~\ref{tab:ancilmodels}, we observe - very much like in the previous section - that the Y-point is closer to the LC, 
 and the luminosity at the LC is lower in case of models using the HCC method. Regarding the dissipation of Poynting flux beyond the LC, the whole series of models (\textbf{Bk})-(\textbf{Br}) display similar and relatively low values of $\Delta L/L_{\rm LC}$, especially if we compare these values to the analogous ones obtained with the LCR method and $\eta=0$ (models \textbf{Bg} and \textbf{Bi}). For most of the phenomenological resistivities considered here (say $\eta\le 10^{-2}$), HCC models are more diffusive than CC models and this global measurement (of luminosity) manifests itself at the local level in a smoother velocity profile along the equatorial current sheet. Contrasting this, local reconnection events with large drift 
 velocities (normal to the ECS) become visible in Fig.~\ref{fig:ETA_COMPARE} (bottom row of panels) for the cases using the CC method (especially in models \textbf{q} and \textbf{r}). At the same time, the width of the resistive layer, in which we can identify an inflow (drift) velocity into the ECS, increases with larger $\eta_{\rm d}$. 

\section{Discussion} 
\label{sec:discussion}

\subsection{Action of non-ideal electric fields in FFE}
\label{sec:nonidealFF}

Our method extracts deviations from the ideal FFE condition $\mathbf{E}\cdot\mathbf{B}=0$ by the algebraic reset 
\begin{align}
	E^i\rightarrow E^k\left(\delta^i_{\hspace{4pt}k}-B_k\frac{B^i}{\mathbf{B}^2}\right).
	\label{eq:DBcutback}
\end{align}
in each sub-step of the time-integrator. This surgical intervention instantaneously achieves (ideal) perpendicularity of electric and magnetic fields. At this point, the results presented in Sect.~\ref{sec:tradesecrets} suggest two different dynamical readjustments. First, a local reconstruction of charge adds an amount of charge into the domain  that can be computed taking the divergence of Eq.~\eqref{eq:DBcutback}. By assuming a pointwise correction as well as $\nabla\cdot\mathbf{B}=0$, one obtains
    \begin{align}
        \rho\rightarrow\rho-\nabla\left[\frac{\mathbf{E}\cdot\mathbf{B}}{|\mathbf{B}|^2}\right]\cdot \mathbf{B}.
        \label{eq:misalignedrecon}
    \end{align}
    
    As there is no discrepancy between the charge density $\rho$ and $\nabla\cdot\mathbf{E}$, the cleaning potential $\Phi$ reduces its role to a true scalar correction of (very small) numerical truncation errors. Non-ideal fields still alter the system of conservation laws by the loss of both, energy conservation, and charge conservation. The charge density is one constituent of the current that acts as a source of Ampère's law. A source or sink of it, due to the addition of charge when fixing the violation of the $\mathbf{E}\cdot\mathbf{B}=0$ constraint, is dynamically relevant and may have a global impact as it can be transported away from the numerical cells where it is initially generated.
    %not a localised artifact.
    
 Second, in a system that transports charge density in a fully conserved way, the correction in Eq.~\eqref{eq:DBcutback} does not induce local alterations of $\rho$. However, it induces a discrepancy $\mathcal{R}$ between $\rho$ and $\nabla\cdot\mathbf{E}$ that corresponds to the same amount that was identified above:
    \begin{align}
        \mathcal{R}\equiv\nabla\cdot\mathbf{E}-\rho\rightarrow\nabla\left[\frac{\mathbf{E}\cdot\mathbf{B}}{|\mathbf{B}|^2}\right]\cdot \mathbf{B}.
        \label{eq:misalignedcons}
    \end{align}

Such a mismatch between the divergence of electric fields and charge density will cause a change of the cleaning potential $\Phi$ that is not necessarily a small correction. Just as the current density, $\Phi$ acts as a source of Ampère's law. Furthermore, $\Phi$ is a correction to its conservative flux. With the same logic as above, it is, thus, dynamically relevant and with a potential impact beyond the places where charge corrections are induced due to the violation of the $\mathbf{E}\cdot\mathbf{B}=0$ constraint. With the similarity of Eqs.~(\ref{eq:misalignedrecon}) and~(\ref{eq:misalignedcons}) it is not surprising that the LCR method may be regarded, in some aspects, as a low $\alpha$ limit of the CC method (with only weak action of the scalar cleaning potential $\Phi$). Strikingly, the action of non-ideal electric fields on a global scale remains relevant even when replacing the algebraic corrections to force-free violations (Eq.~\ref{eq:DBcutback}) by the continuous action of a driving current as introduced in Eq.~(\ref{eq:FFResCurrentPerpendicular}). In this case, the non-ideal term
\begin{align}
    \mathcal{S}_{\rm ni}=\kappa_I\frac{\mathbf{E}\cdot\mathbf{B}}{|\mathbf{B}|^2}\mathbf{B}=\kappa_I\mathbf{E}_\parallel
    \label{eq:nonidealsource}
\end{align}
emerges a source of heating in the energy evolution equation \citep[][Eq.~2.73]{MahlmannPhD}. Its purpose is to continuously drive the electromagnetic field to a force-free state. The results presented in Sect.~\ref{sec:drivingfocus} allow two notable interpretations: i) the diffusion induced by Eq.~(\ref{eq:nonidealsource}) does not significantly change the dissipation of Poynting flux along the ECS (measured by $\Delta L/L_{\rm LC}$ obtained by applying Eq.~\ref{eq:DBcutback}); and ii) explicit driving currents on cell-centered meshes are not able to overcome the need of charge conservation in FFE. All the conducted simulations (including those employing driving currents) always reduce to the most diffusive limit whenever the LCR method is applied.

\subsection{Time-scales of the hyperbolic/parabolic cleaning}
\label{sec:cleaningscales}

Equations \eqref{eq:Efield} and \eqref{eq:Phi} can be manipulated to show that both, the scalar potential $\Phi$ and the difference parameter $\mathcal{R}$, obey the telegraph equation \citep{Komissarov2007MNRAS.382..995}
\begin{align}
    &\partial_{tt}^2 \Phi + \kappa_\Phi \partial_t \Phi - c_\Phi^2\nabla^2 \Phi=0 \label{eq:telegraph1}\\
    &\partial_{tt}^2 \mathcal{R} + \kappa_\Phi \partial_t \mathcal{R} - c_\Phi^2\nabla^2 \mathcal{R}=0. 
\end{align}
Let $\tau$ and $l$ be the characteristic time and length scales of change of $\Phi$ (or $\mathcal{R}$), respectively. From Eq.~\eqref{eq:telegraph1}, approximating $\partial_t\Phi\approx \Phi/\tau$ and $\nabla\Phi\approx \Phi/l$, we obtain
\begin{align}
    &\frac{\Phi}{\tau^2} + \kappa_\Phi \frac{\Phi}{\tau} - c_\Phi^2\frac{\Phi}{l^2}\approx 0 . \label{eq:telegraph2}
\end{align}
In the limit $\tau\ll 1/\kappa_\Phi$, Eq.~\eqref{eq:telegraph2} yields $\tau\approx l/c_\Phi\equiv\tau_a$, where $\tau_a$ has the meaning of an advection timescale for cleaning errors. In the complementary limit $\tau\gg 1/\kappa_\Phi$, one obtains  $\tau\approx l^2 \kappa_\Phi/c_\Phi^2\equiv\tau_d$, where $\tau_d$ can be interpreted as the diffusion timescale for the cleaning of errors. The ratio between both time scales is precisely the parameter $\alpha$ defined in Eq.\,\eqref{eq:alphadimensionless}, i.e. $\alpha = \tau_d/\tau_a$.    

As has been noted throughout the literature \citep[e.g.][]{Mignone2010,Mahlmann2019,Mahlmann2020c}, mostly in the context of cleaning errors to the $\nabla\cdot\mathbf{B}=0$ constraint, a careful calibration of the parameters controlling hyperbolic/parabolic cleaning is necessary. The same holds true for the cleaning of errors to the $\rho=\nabla\cdot\mathbf{E}$ constraint, and we conducted a thorough calibration for the simulations shown in this paper. One way of optimizing the cleaning parameter $\alpha$ is to evaluate the dissipation induced by the source terms to the energy evolution equation, as it was presented in \citet[][Eq.~2.73]{MahlmannPhD}. The relevant channels are dissipation by cleaning ($\propto c_\Phi^2\mathbf{E}\cdot\nabla\Phi$), as well as Ohmic heating fueled by non-ideal electric fields ($\propto\mathbf{E}\cdot\mathbf{J}$). 

With the results in Fig.~\ref{fig:QDissipation} we find that employing the $\Phi$ cleaning with $\alpha\approx 1$ yields optimal results. This means that an optimal regime is reached when the timescales of dissipation ($\tau_d$) and advection ($\tau_a$) of numerical errors induced by the violation of FFE constraints (in the aligned force-free rotator, predominantly at the ECS) are of the same order. First, the dissipation by corrections of non-ideal electric fields is minimised across all models in this regime (light blue stripe). Second, the dissipation by the cleaning potential stabilises at a steady value across resolutions. This plateau roughly coincides with the blue shaded region and can be interpreted as a trade-off between excessively weak cleaning (inducing larger dissipation for small values of $\alpha$ because of the oscillatory behavior of the electric field after applying corrections to enforce the force-free regime) and the extreme case of over-damping that drives the scalar function $\Phi$ to zero very rapidly. 

\subsection{Magnetospheric structure}
\label{sec:structure}
\begin{figure}
  \centering
  \includegraphics[width=0.47\textwidth]{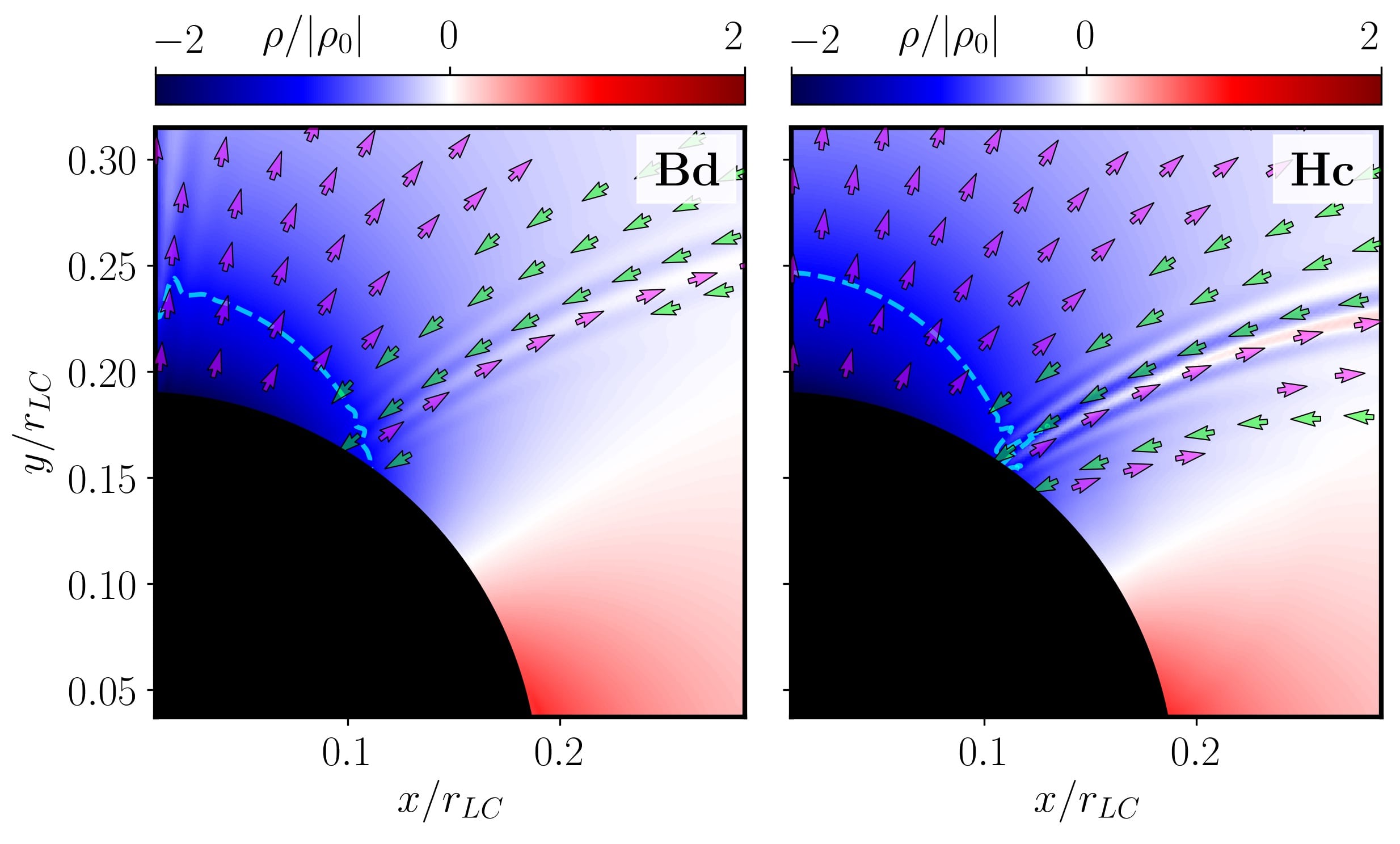}\\
  \includegraphics[width=0.47\textwidth]{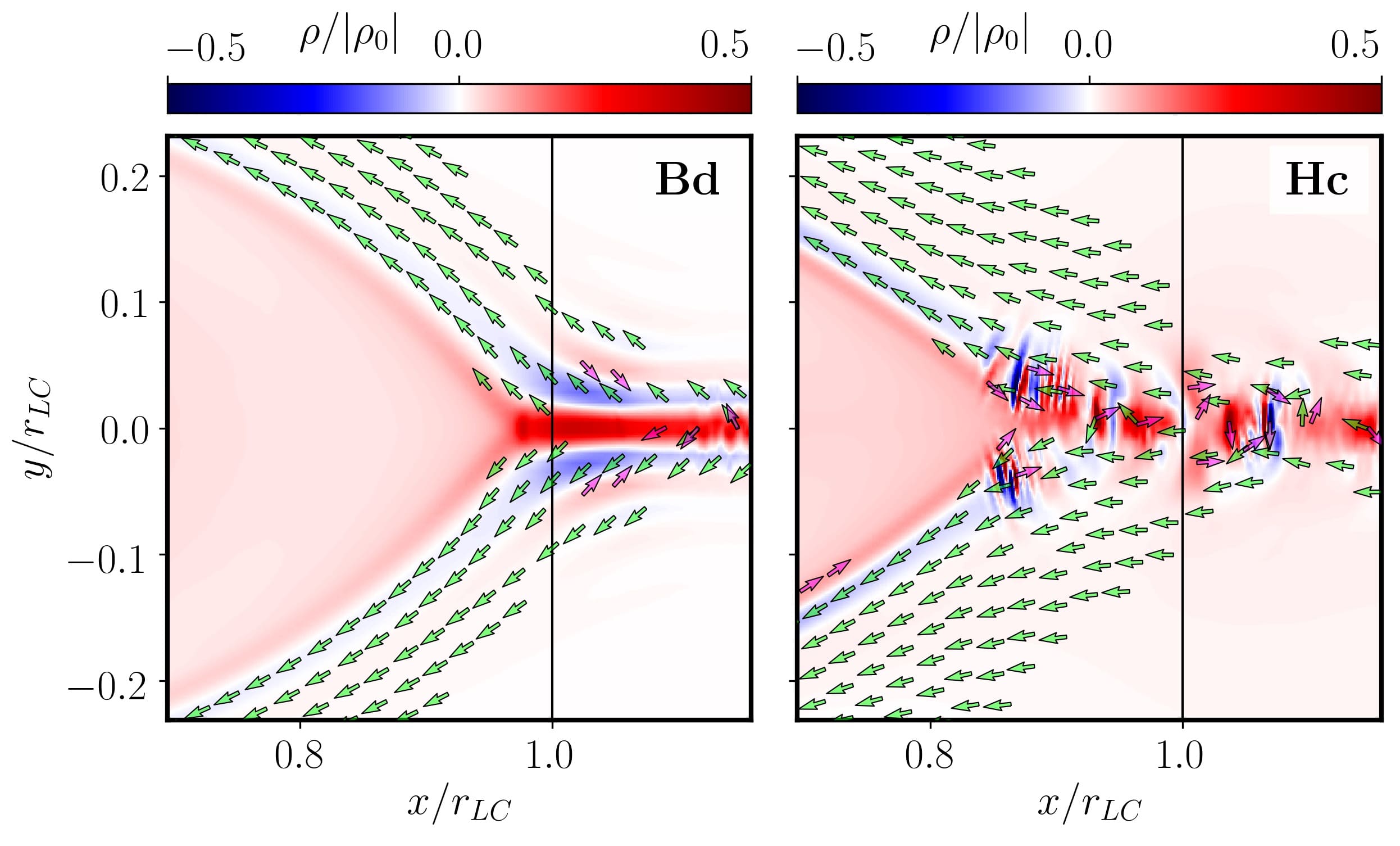}\\
  \includegraphics[width=0.47\textwidth]{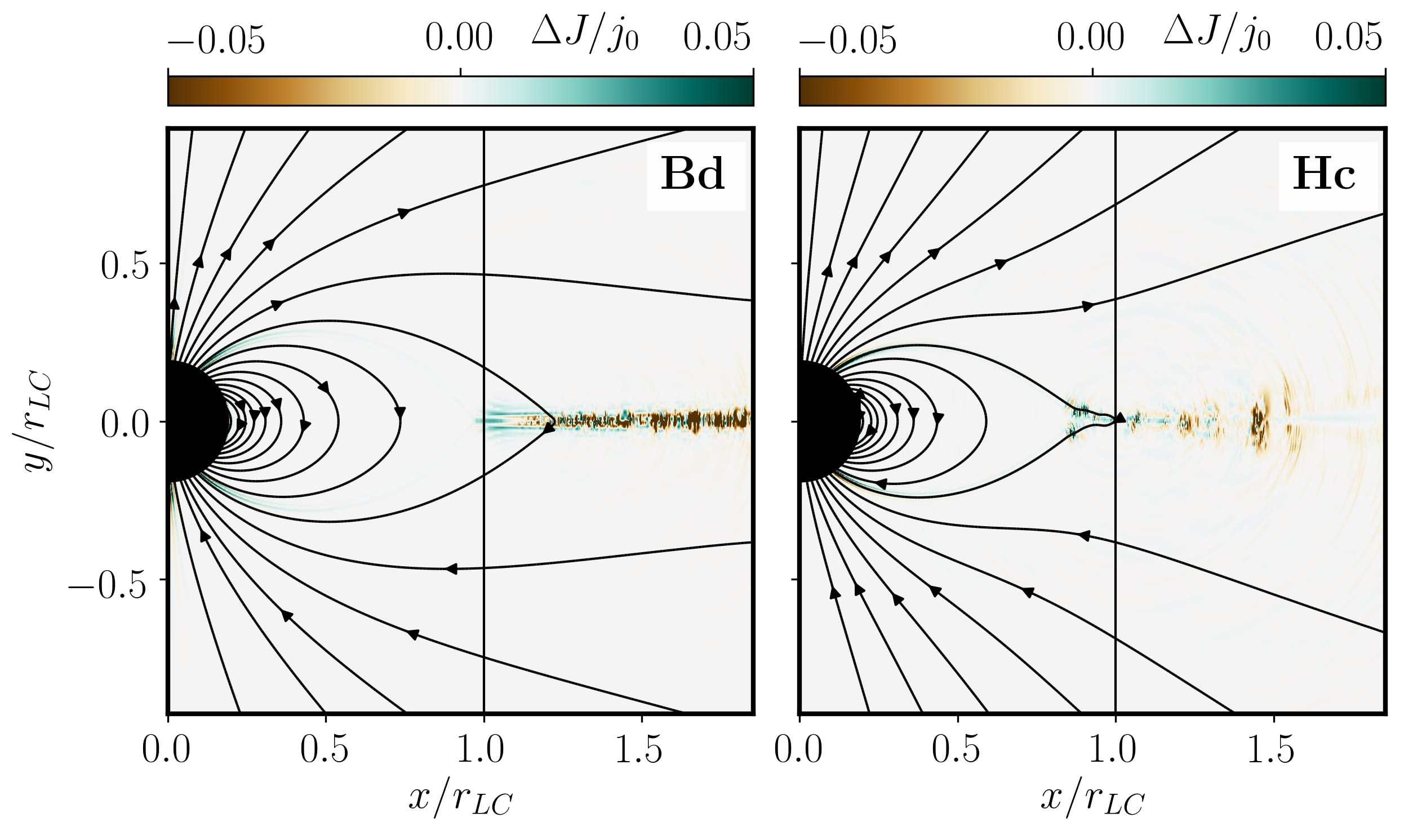}
  \vspace{-6pt}
  \caption{Charge and current structure in time-dependent aligned rotator models. Upper and mid panels: Charge ($\rho$; colored background, normalised to the absolute value of the Goldreich-Julian charge density, $|\rho_0|$) and current ($\mathbf{j}_\parallel$; arrows, normalised to $\rho_0 c$) structure of selected models (see panel legends). We visualise two different zooms into the magnetosphere, namely the polar cap region (top panels), and the Y-point vicinity (bottom panels). The direction of the current flow is indicated by arrows for regions where $j_\parallel/j_0>0.01$. The cyan dashed line denotes the location where $\rho=\rho_0$ (specifically, in between that line and the stellar surface, the charge density is larger than the Goldreich-Julian charge density). Bottom panels: Difference between the current $\mathbf{j}$ in our time-dependent models and the current in the steady state case  $\mathbf{j}_{\rm steady}=\nabla\times\mathbf{B}$, normalised to the Goldreich-Julian current density $j_0$. We display the poloidal magnetic field lines; the vertical black line denotes the position of the LC.}
\label{fig:CURRENT_ZOOM}
\end{figure}

In Fig.~\ref{fig:CURRENT_ZOOM}, we display the structure of currents and charges in two selected regions of the magnetosphere. Specifically, we examine a location close to the polar cap and another one around the Y-point for models using the LCR method (model \textbf{Bd}) and CC method (model \textbf{Hc}). Both models have the same value of $\alpha=0.7$ and the same numerical resolution. The overall charge distribution looks qualitatively like the one obtained by previous force-free models \citep[e.g.][in axial symmetry or \citealt{Kalapotharakos_2009A&A...496..495, Kalapotharakos_2012ApJ...749....2} in three dimensions]{Parfrey_2012MNRAS.423.1416} and in a number of PIC simulations of aligned rotators \citep[e.g.,][]{Chen_2014ApJ...795L..22,Philippov2014,Cerutti2016,Brambilla_2018ApJ...858...81}. Negative charges fill the regions above the polar caps, while positive ones fill the closed magnetosphere up to the Y-point. The structure of the Y-point and of the ECS adjacent to it consists of a set of charge layers of alternating sign stacked vertically in model \textbf{Bd} (Fig.~\ref{fig:CURRENT_ZOOM}, bottom left panel). This layered structure is also observed in, e.g. \citet[][cf. their Fig.~16]{Parfrey_2012MNRAS.423.1416}. Along the equatorial region, a positively charged layer with a thickness $\sim 0.027r_\LC$ emerges, reproducing the results of a positive surface charge density along the ECS shown in \cite{Timokhin2006}. In model \textbf{Hc}, the layered charge structure is destroyed due to time-variable episodes of reconnection along the ECS. Still, the positively charged central layer is intertwined with regions of negative charge. 
The (more) variable structure of the ECS in models using the CC method arises because of the existence of a finite time induced by restoring the stationary condition $\nabla\cdot\mathbf{E}=\rho$. This finite time is brought about by the (finite) propagation speed ($c_\Phi\sim c$) of the hyperbolic part of the equation controlling $\Phi$, and modulated by the (finite) diffusion timescale $\tau_d$ (see Sect.~\ref{sec:cleaningscales}). In the LCR method, this time is zero, as to say the restoration of $\nabla\cdot\mathbf{E}=\rho$ is instantaneous. However, the current does not follow the change in the charge density distribution instantaneously. 

The directions of the poloidal current (displayed by the colored arrows in Fig.~\ref{fig:CURRENT_ZOOM}) shows a return current flowing along the ECS and continuing over the closed zone separatrix converging on the Y-point and extending up to the stellar surface. However, the structure of the current is not simple. Among currents flowing towards the stellar surface, we also find anti-parallel currents flowing away from the surface. We point out the interesting observation that above the polar caps a region with super Goldreigh-Julian charge density, i.e. $\rho>\rho_0$ (enclosed by the cyan dashed line in the upper panels of Fig.~\ref{fig:CURRENT_ZOOM}), emerges.

Under the assumption of stationarity, in a small layer around the ECS with a thickness $2h\ll r$, \cite{Contopoulos_2014ApJ...781...46} constructed a simple analytic model for the electromagnetic structure, where the electric field toroidal component is zero. However, in our 
models, we find a time-dependent $|E_\phi(r,z,t)|\ne 0$ in places along the ECS (and $x>1$) where the magnetic dominance condition is breached. Indeed, the model using the CC method shows a quasi-periodic pattern of the form  $E_\phi(r,z,t)\sim E_\phi(r,h)\sin\left[8\pi(x-t)\right] $. 
For the model using the LCR method, the spatial frequency is about two times larger. The other two components of the electric field roughly follow the analytic model of \cite{Contopoulos_2014ApJ...781...46}, namely, they follow the relations $E_r(x,h)\approx -xB_z(x,z)$ and $E_z(x,h)\approx xB_r(x,z)$. Because of the presence of displacement currents and the (comparatively much smaller) contributions of the term $c_\Phi^2\nabla\Phi$ in Eq.~\eqref{eq:Efield}, the current density in the ECS is not given by $\mathbf{j}_{\rm steady}=\nabla\times\mathbf{B}$ as in the stationary analytic model of \cite{Contopoulos_2014ApJ...781...46}. This is lucidly represented in the bottom panels of Fig.~\ref{fig:CURRENT_ZOOM}, where the ratio $\Delta J/j_0=(|\mathbf{j}|-|\nabla\times\mathbf{B}|)/j_{0}$ differs from zero (indicating that the current density includes contributions other than $\mathbf{j}_{\rm steady}$) mostly in the vicinity of the ECS and also along the current sheets surrounding the closed magnetospheric region. 
At this point, there is an interesting difference between models using LCR and CC methods, namely, models with a local charge reconstruction have a current smaller than $\mathbf{j}_{\rm steady}$ ($\Delta J/j_0<0$) along most of the current sheet. In contrast, the model \textbf{Hc} shows smaller average deviations along the ECS between $\mathbf{j}$ and $\mathbf{j}_{\rm steady}$, and surrounding it in intermediate patches beyond the LC. That finding is unexpected given the larger variability along the ECS encountered, in general, for CC models.

\subsection{Diffusivity in force-free magnetospheres}
\label{sec:diffusivitydiscuss}

\begin{figure*}
  \centering
  \includegraphics[width=1.0\textwidth]{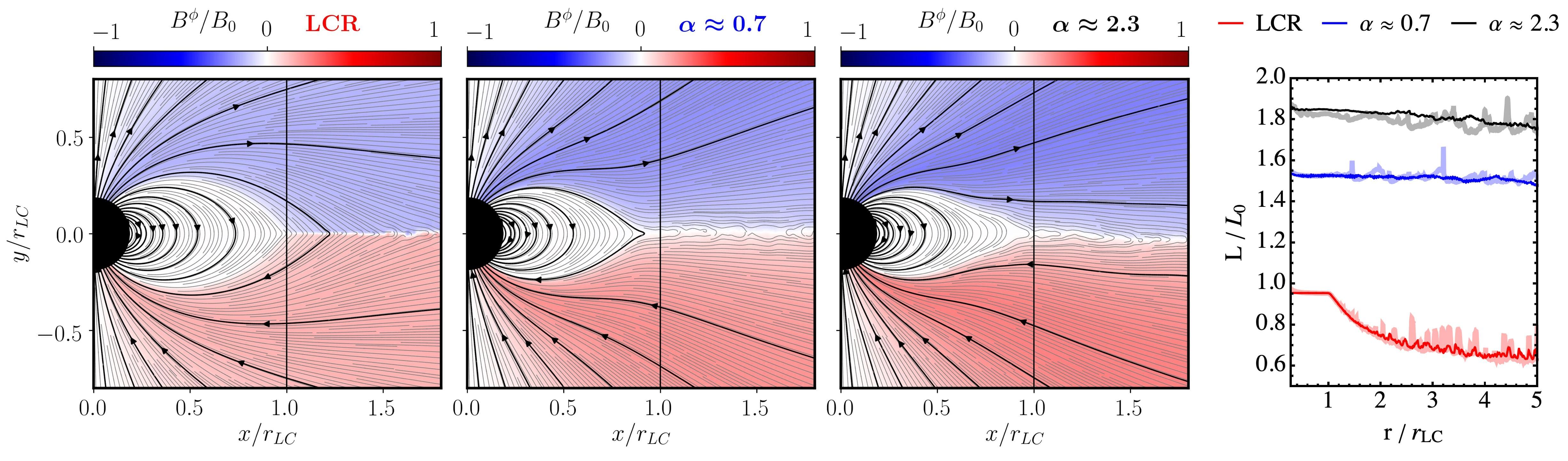}
  \vspace{-18pt}
  \caption{Averaging of magnetospheric fields and Poynting fluxes during a time of $\Delta t=(7.5-6.8)t_{\rm p}=1.3t_{\rm p}$ for selected models ($\mathbf{Ba}$, $\mathbf{Be}$, and $\mathbf{Bb}$). Small variations beyond the light cylinder, especially those observed in Fig.~\ref{fig:Poynting} (shown again as thick, transparent lines in the background of the right panel) are compensated across timescale $\Delta t\gtrsim 1t_{\rm p}$.}
\label{fig:Averaging}
\end{figure*}

Resetting the charge density $\rho$ to the instantaneous value of $\nabla\cdot\mathbf{E}$ assures compatibility of electric fields and the corresponding charge distribution. However, removing significant fractions of the electric field from the domain for ensuring FFE conditions alters the charge distribution in the domain by an amount that is not necessarily of the order of the truncation error. 
In our default (CC) scheme \citep{Mahlmann2020b}, the charge is evolved with a separate continuity equation. Changes to the electric field by the algebraic reset of force-free errors introduce a misalignment between the charge density and the divergence of electric fields. The degree of localization of this misalignment to inherently non force-free regions, such as current sheets, can be interpreted as a diffusive length scale. We control this localization in our hyperbolic/parabolic cleaning procedure with the parameter $\alpha$, that expresses the ratio of diffusion to advection timescales for the cleaning errors.

The LCR method turns out to be the most diffusive one, using as a measurement of the diffusivity the dissipation of Poynting flux beyond the LC. In the models resorting to the LCR method, the effect of violations to ideal FFE conditions spreads throughout the domain, with notable consequences for the global energetics. While the Y-point relaxes to a location close to the LC (cf. Fig.~\ref{fig:FFCOND_COMPARE}), field lines reconnect across the ECS and induce a notable dissipation of the outward transported energy ($\sim 30\%$).  When augmented by a conservative evolution of the charge density with a suitable calibration of the divergence cleaning parameter $\alpha$, reconnection beyond the LC is greatly reduced - as is the dissipation of energy. At the same time, the Y-point is pushed away from the light cylinder (cf. Fig.~\ref{fig:FFCOND_COMPARE}), and the overall luminosity increases. Though the action of the cleaning potential $\Phi$ can become excessively large for $\alpha\gg 1$, the increase of the luminosity with the corresponding change of the Y-point location roughly follows the trend that is theoretically expected for small and intermediate values of $\alpha$ (cf. Fig.~\ref{fig:YpointLuminosity}).

We explore an interpretation of our results in light of the parameterization of the diffusivity beyond the LC in terms of the pair formation multiplicity $\kappa$ suggested by \cite{Contopoulos2020}. These authors show that smaller values of $\kappa$ eventuate in more dissipation at the ECS and, conversely, $\kappa\gg 1$ yields very little or no dissipation at all. When drawing a direct comparison to the results presented throughout this paper, the models endowed with the LCR method would correspond to values of $\kappa\sim 1$. Models employing our the CC scheme (our default), would compare to values $\kappa\gg 1$. The physical conditions in a typical pulsar magnetosphere tend to produce many pairs per Goldreich-Julian charge particle in the polar cap \citep[e.g.][and references therein]{Timokhin_2015ApJ...810..144,Contopoulos_2019MNRAS.482L..50}. Hence, in this regard, our CC models potentially reproduce the usual conditions met in actual rotating pulsars more closely, though the limitations set by the force-free regime naturally persist.

The panels of Fig.~\ref{fig:FFCOND_COMPARE} that show the location of violations of the FFE constraints illustrate two notable aspects: i) deviations from $\mathbf{E}\cdot\mathbf{B}=0$ are not a localised phenomenon. Emerging from genuinely non-ideal regions, such as current sheets, non-ideal fields can spread throughout the domain. ii) A change in the treatment of charge conservation not only minimises the dissipation induced by corrections of the $\mathbf{E}\cdot\mathbf{B}=0$ condition, but also changes the electromagnetic structure of the current sheet itself. Cases $\mathbf{Ba}/\mathbf{Bd}$ show a region with $\mathbf{E}^2-\mathbf{B}^2>0$ along the length of the current sheet. Contrasting this, such violations only occur at X-points in the current sheet of models $\mathbf{Bb}/\mathbf{Be}$. The acceleration of particles and the production of radiation demand the existence of an electric field component parallel to the magnetic field, namely that $\mathbf{E}\cdot\mathbf{B}\ne 0$. Hence, the structure of the regions of the magnetosphere, where the strongest violations of the $\mathbf{E}\cdot\mathbf{B}=0$ condition occur, are likely closely related to the production of pulsar radiation \citep[e.g.][]{Timokhin_2013MNRAS.429...20}. Looking at Fig.~\ref{fig:FFCOND_COMPARE}, these violations, although extended in the magnetosphere as stated above, are maximised in the vicinity of the Y-point, suggesting that the Y-point is the most important site for particle acceleration in the magnetosphere. This result is backed up by PIC simulations of, e.g. \cite{Chen_2014ApJ...795L..22} and gives support to the theoretical "ring-of-fire" model of \cite{Contopoulos_Stefanou_2019MNRAS.487..952}.

Violations to the ideal force-free conditions in combination with the transport of charge conservation errors originating during their correction (Sect.~\ref{sec:nonidealFF}) are the main driver of diffusivity in force-free aligned pulsar magnetospheres. The relative importance of the dissipation triggered by the enforcement of either the  $\mathbf{E}\cdot\mathbf{B}=0$ or the $\mathbf{E}^2-\mathbf{B}^2<0$ conditions is similar across these two particular channels, as we can observe in the magnitudes of the electric energy lost within a given time step in Fig.~\ref{fig:QDissipation}. However, the magnitude of the dissipation triggered by each of them may depend on the order and frequency with which they are applied, as well as on the mesh and time-integrator \citep[cf.][]{Spitkovsky2006}. 
Nevertheless, the dominant contribution to the dissipation along these channels stems from the ECS, where the magnetic dominance condition is chronically breached. We suggest that the magnetic dominance condition is really the origin of the differences between the various methods of dealing with the charge treatment as presented in this paper. In the HCC models using local charge reconstruction only in grid zones where $\mathbf{E}^2-\mathbf{B}^2>0$, the dissipation through this channel is minimised (see light green colored squares in the upper panel of Fig.~\ref{fig:QDissipation}), and the Ohmic dissipation is maximised (mid-panel of Fig.~\ref{fig:QDissipation}).
Since $\mathbf{E}^2-\mathbf{B}^2>0$ is only reached at the ECS, it is standing to reason that the specific restoration of the magnetic dominance constraint may induce global changes in the magnetospheric structure, its luminosity, and, certainly, on the amount of dissipation of Poynting flux beyond the LC.

The resistivity models used beyond the LC in Sect.~\ref{sec:diffusivityfocus} provide twofold insight. First, they allow us to estimate the numerically induced diffusivity $\eta_0$ 
across the ECS. The effect of $\eta_d$ will only become noticeable when the phenomenological resistivity is larger than the numerical diffusivity of the method. From Fig.~\ref{fig:ETA_COMPARE}
%and~\ref{fig:ETA_COMPARE_Q10} 
we can estimate that $\eta_0\lesssim 10^{-1}$. Second, and in line with the results presented in \citet{Mahlmann2020c}, a choice of $\eta_d\gtrsim 10^{-1}$ in the current presented in Eq.~(\ref{eq:FFResCurrentPerpendicular}) allows to properly model dynamics of the resistive layer of the ECS. Increasing $\eta_d$ gradually drives relatively large scale
inflows into the ECS (as traced by the drift velocity component perpendicular to it), effectively mimicking physical dynamics in the non-ideal region. As we established very competitive convergence of our high-order FFE method \citep{Mahlmann2020c}, we suggest that this relatively large value of $\eta_0$ is, indeed, induced by the non-ideal fields emerging in the ECS. In this context it becomes clear why several FFE methods need to employ special treatments of the ECS in the pulsar magnetosphere \citep{McKinney2006,Etienne2017}, namely, to reduce the extent of the diffusive regions by an ad-hoc prescription.

As stated in, e.g. \cite{Timokhin_2013MNRAS.429...20}, a necessary criterion imposed by observational constraints on the pulsar magnetosphere is the \emph{stationarity} in a statistical sense. In other words, any local fluctuations on timescales smaller than the LC light-crossing time $\tau_{\rm LC}=r_{\rm LC}/c$ (or more likely, over the rotational period of the pulsar $t_{\rm p}$) that average to a stationary state, may also account for the stability of the pulsar mean profiles and sharpness of the peaks in the spectra of gamma-ray pulsars. A very salient feature related to the treatment of the charge conservation equation in FFE is the obvious time-dependence of the magnetosphere, driven by episodes of magnetic reconnection along the ECS. However, the magnetospheres resulting from the CC method (including a suitable conservative treatment of the charge) are stationary if we average them out over timescales comparable to $\tau_{\rm LC}$. In order to support this statement, we show the time-averaged map of the toroidal magnetic field  over an interval $\Delta t$ slightly larger than one pulsar rotational period in Fig.~\ref{fig:Averaging} (left panels). We find even stronger evidence than in the polar distribution of the toroidal field when evaluating the Poynting flux as a function of distance for the averaged models. It displays a radial dependence which radically smooths the spatial variability (Fig.~\ref{fig:Averaging}, right panel). We further probed the long-term stability of the representative models $\mathbf{Ba}$, $\mathbf{Be}$, and $\mathbf{Bb}$ by tracking their evolution during $\gtrsim 30$ rotational periods \citep{SupplementaryMediaA}. The models show stability over such time-scales in all the characteristic properties discussed in the previous section, especially regarding the Y-point location and pulsar luminosity.

\section{Conclusions} 
\label{sec:conclusion}

In a deep exploration with our recently developed force-free code \citep{Mahlmann2020b,Mahlmann2020c}, we exploit the diffusion time-scale induced by hyperbolic/parabolic cleaning of charge conservation errors (Sect.~\ref{sec:cleaningscales}) to quantify an aspect that has not been systematically assessed  so far in many FFE simulations, including our own. Namely, the \emph{global} imprint of \emph{local} violations to the force-free constraints. At the example of the force-free aligned rotator magnetosphere, we demonstrate that balancing the amount of damping and advection of charge conservation errors (encoded in the parameter $\alpha$) can alter the global structure of the simulated magnetosphere. Specifically, by decreasing the amount of numerical diffusion arising from violations to the FFE constraints, the Y-point moves away from the light cylinder while the outgoing Poynting flux increases by a factor of a few (Sect.~\ref{sec:forcefreealigned}). 

In summary, our exploration clarifies several \emph{technical} aspects that should become central for the assessment of (global) FFE simulations. First, the localization of force-free violations to small regions in resistive layers, such as current sheets, are key to reduce the diffusivity that is induced into FFE by non-ideal electric fields or breaches of magnetic dominance (Sect.~\ref{sec:nonidealFF}). In our method, we achieve and control such a localization by combining a conservative evolution of charge density with a hyperbolic/parabolic cleaning of errors to Gauss' law. We suggest that $\alpha\lesssim 1$ is an optimal parameter for the minimization of the combined channels of numerical diffusion (Sect.~\ref{sec:focuseddominance}). Second, in the inherently non-ideal aligned rotator magnetosphere, the ECS is the 
%one 
main source of numerical diffusion by inducing strong field gradients and violations to the FFE constraints. We identify a strong dependence of the dynamics on the specific treatment of violations to the magnetic dominance condition (Sect.~\ref{sec:diffusivitydiscuss}). Third, the extreme nature of algebraic corrections to the force-free conditions are not the main reason for the luminosity dependence of the global magnetosphere. Driving currents yield very similar results (Sect.~\ref{sec:drivingfocus}).
Finally, different treatments of force-free violations, especially at Y-points and current sheets, are likely to change the resistive time scales of the evolution and to have a notable impact on the equilibrium magnetospheres. We extend our analysis to so-called phenomenological resistivity models, where adapted driving currents mimic the development of resistive layers around current sheets (Sect.~\ref{sec:diffusivityfocus}).

\textit{FFE is a robust way to model energy flows in highly magnetised plasma}, as we find in the magnetospheres of many astrophysical objects. This statement can safely be extended to situations in which the global dynamics of field lines drives the transient appearance of inherently non-ideal regions, such as current sheets. Specifically, we argued this in the context of accretion of magnetic loops onto rapidly spinning BHs \citep{Mahlmann2020}, and the shearing of fields driven by interacting Alfvén waves \citep{Ripperda2021}. However, in situations where areas of genuine (physical) resistivity drive the global field line dynamics, employing FFE methods has to be carefully benchmarked. This work is, to our knowledge, the first extensive calibration of an FFE method for the specific application of astrophysical magnetospheres. We find the global flows of energy to be extremely sensitive to the treatment of FFE violations. Our results agree with comparable numerical surveys throughout the literature only in the limit of strong numerical diffusion induced by the ECS. 

Finally, we suggest that scenarios like the aligned rotator should be handled with care when used as a standard test for FFE methods. The stability and magnitude of the Poynting flux beyond the light cylinder can be used as a primer for the diffusivity of the respective method. It could be argued that operating in the well-established, but ultimately limited, regimes of \emph{ideal} fluid approximations for the modeling of global scenarios that are affected or even driven by genuinely resistive effects will require more and more care in the future. The desire to overcome many orders of scale separation has to go hand-in-hand with a deep understanding of the diffusive properties of the employed numerical methods. Until we can transition into an era of multi-regime astro-plasma codes, we find it reassuring to have the limits of our FFE method laid out transparently.

\section*{Acknowledgements}

This work has been supported by the Spanish Ministry of Science, Education and Universities (PGC2018-095984-B-I00) and the Valencian Community (PROMETEU/2019/071). We furthermore thank for support from the COST Actions PHAROS CA16214 and GWverse CA16104. This manuscript relies vastly on high performance computing resources. They were provided with the \textit{LluisVives} machine at the Servei d’Informàtica of the \textit{Universitat de València} (financed by the FEDER funds for Scientific Infrastructures; IDIFEDER-2018-063) and extensively supplemented by allocations on the \textit{MareNostrum} and \textit{Tirant} supercomputers of the \textit{Spanish Supercomputing Network} (AECT-2021-1-0006, AECT-2021-1-0007).

\section*{Data Availability}

The data underlying this article will be shared on reasonable request to the corresponding author.

%%%%%%%%%%%%%%%%%%%% REFERENCES %%%%%%%%%%%%%%%%%%

\bibliographystyle{mnras}
\bibliography{literature.bib} 

%%%%%%%%%%%%%%%%%%%%%%%%%%%%%%%%%%%%%%%%%%%%%%%%%%

%%%%%%%%%%%%%%%%% APPENDICES %%%%%%%%%%%%%%%%%%%%%

\appendix

% Don't change these lines
\bsp	% typesetting comment
\label{lastpage}
\end{document}